\documentclass[%
 twocolumn,
superscriptaddress,
amsmath,amssymb,
aps,
prb,
citeautoscript,
]{revtex4-1}

\usepackage{subfigure}
\usepackage{comment}
\usepackage{color}
\usepackage{physics}
\usepackage{graphicx}
\usepackage{dcolumn}
\usepackage{bm}

\begin{document}

\title{Topological ordering in the Majorana Toric Code}

\author{Alexander Ziesen}\email{alexander.ziesen@rwth-aachen.de} 
\affiliation{JARA Institute for Quantum Information, RWTH Aachen University}
\author{Fabian Hassler} 
\affiliation{JARA Institute for Quantum Information, RWTH Aachen University}
\author{Ananda Roy}%
\affiliation{Department of Physics, T42, Technische Universit\"at M\"unchen, 85748 Garching, Germany}%

\date{12 August 2019}

\begin{abstract}
At zero temperature, a two-dimensional lattice of Majorana zero modes on mesoscopic superconducting islands exhibits a topologically-ordered toric code phase. Recently, a Landau field theory was used to describe the different phases of the aforementioned system and the phase-transitions separating them. While the field theory provides details on the properties of the system close to the phase-transitions, signatures of topological ordering in the different phases have not been computed. This is the primary goal of the current work. We describe a lattice gauge theory of the Majorana toric code in terms of $\mathrm{U}(1)$ matter fields coupled to an emergent $\mathbb{Z}_2$ gauge field. Subsequently, we use a generalized Wilson-loop order-parameter, the equal-time Fredenhagen-Marcu order parameter, to characterize the topological ordering in the different phases. Our computation provides evidence of the toric code phase both in the Mott insulator and the charge-$2e$ superconductor phases, while showing that the toric code phase disappears in the charge-$e$ superconductor phase. In addition, we perturbatively analyze the influence of Cooper pair tunneling on the topological gap of the toric code in the limit of strong charging energy and show that the toric code phase is, in fact, stabilized by the Cooper pair tunneling. Our results are relevant for experimental realizations of the Majorana toric code. 
\end{abstract}

\maketitle

\section{\label{sec:intro}Introduction}
A promising candidate for realizing fault-tolerant quantum computation is Kitaev's two-dimensional toric code~\cite{ToricCode, ToricCode_anyons, KITAEV20062,Fowler_TC}. The ground (code) space is topologically ordered~\cite{ToricCode, Wen_string_net, Wen_TO_in_GS}
and is four-fold degenerate. Thus, it can encode two logical qubits. These encoded logical qubits are robust to local perturbations since the degeneracy of the ground space depends only on the topology of the embedding space where the code is implemented. There are several approaches to realize the toric code. The first approach involves tessellating a two-dimensional plane with a regular lattice, with physical qubits on the links of the lattice. The Hamiltonian of the system is the toric code Hamiltonian which comprises vertex terms and plaquette terms. Each vertex term is a product of $\sigma_x$ operators of the qubits residing on the links incident at the vertex, while each plaquette term is a product of $\sigma_z$ operators of the qubits residing on the links around a plaquette~\cite{ToricCode_anyons}. A different approach to realize the toric code is to design Hamiltonians of interacting many-body systems which, in the low-energy sector, give rise to the toric code Hamiltonian. This is the case of Kitaev's honeycomb model, which describes SU(2) spins, interacting with alternating $\sigma_x\sigma_x$, $\sigma_y \sigma_y$ and $\sigma_z\sigma_z$ interactions around a plaquette~\cite{KITAEV20062}. In this work, we concentrate on a different implementation of the latter approach involving a lattice of mesoscopic superconducting islands with Majorana zero modes (MZM-s) on each island~\cite{XuFu, Terhal2012, Nussikov, Vijay_MZM, LandauPlugge, Karzig_MZM, Litinski_MZM, Burello_2D_different_system,Lit_von_Oppen_Fermion_Codes, Wille_networks,Plugge_Flensberg_universal_MBQ, Hoffmann_Majo_spin}. 
Note that any physical implementation of either approaches eventually also involves active  measurements which evacuate the entropy due to thermal fluctuations since the topological ordering of the toric code disappears at any nonzero temperature. A toric code built out of MZM-based physical qubits holds the promise of better error correction properties compared to its superconducting transmon qubit-based counterpart. This is due to the potentially superior coherence properties of the MZM-based physical qubits~\cite{Plugge_MZM_arch_advantage,LandauPlugge, Vijay_2016_TC_MZM_arch_adv, Vijay_MZM, Karzig_MZM}. 

The physical system of the Majorana toric code comprises mesoscopic superconducting islands which have finite charging energy ($E_C$). These nearest-neighboring islands are separated by tunnel junctions (see below for details), through which Cooper-pairs can tunnel coherently between islands (at rate $E_J$). Furthermore, due to the presence of the MZM-s, single electrons can also coherently tunnel between neighboring islands (at rate $E_M$). It was shown that for $E_J\ll E_M\ll E_C$, the system is in a topologically ordered toric code phase, while being a Mott-insulator~\cite{XuFu, LandauPlugge}. The presence of the MZM-s lead to an emergent $\mathbb{Z}_2$ gauge field and the topological ordering of the toric code. Increase of $E_M/E_C$, while keeping $E_J$ much smaller than $E_M, E_C$, causes the system to undergo a 3D-XY type quantum phase transition to a charge-$e$ superconductor phase, the latter property arising from the dominant tunneling of single electrons. In this phase, the topological ordering of the toric code disappears~\cite{XuFu}. On the other hand, for $E_M\ll E_C\ll E_J$, the system was again shown to be in a topologically ordered toric code phase, while being a charge-$2e$ superconductor~\cite{Terhal2012}, the latter property arising from the dominant tunneling of Cooper-pairs. Increase of $E_M/E_C$, keeping $E_J$ significantly larger than $E_M, E_C$, causes the system to undergo a 3D-Ising type quantum phase-transition to a charge-$e$ superconductor phase (without toric code ordering). These two different limiting cases were shown to be connected by tricritical points and first order transitions~\cite{RoyTerhal2017}. Using a Landau-Ginzburg-Wilson field theory, the phase-diagram of the model for general $E_C, E_M, E_J$ was predicted~\cite{RoyTerhal2017} and experimentally accessible charge transport characteristics were also computed~\cite{Roy2017}. However, the above field theory calculations focused on the charge characteristics of the system and provided only indirect evidence of the existence of the topological order.

Our goal, in this work, is to fill this void and compute non-local order parameters which provide direct evidence of the presence/absence of topological ordering in the different phases. In contrast to the coarse-grained field theory, we analyze the system at a microsopic lattice level, by transforming the problem to a lattice gauge theory one with $\mathrm{U}(1)$ matter fields and a $\mathbb{Z}_2$ gauge field. In this mapping the emergent $\mathbb{Z}_2$ gauge field of the toric code arises naturally. The coupling between gauge and matter fields causes additional complications in the detection of topological ordering. For a gauge theory which has local symmetry, in absence of matter fields, the Wilson loop is the relevant {\it non-local} order parameter~\cite{Elitzur_1975, KogutIntro}. The characteristic area vs perimeter law decay of the latter is able to distinguish between the different phases of the theory. This should be contrasted to a theory with global symmetry, where the phases are distinguished by a {\it local} order parameter. However, the presence of matter fields in a gauge theory changes the situation since the fluctuations of the matter field can screen the interaction between the charges of the gauge field. This can cause the Wilson loop to decay with perimeter law in all the phases~\cite{FradShenk}. As will be shown in this work, this is indeed the case for the Majorana toric code. For such theories, the concept of Wilson loops has to be generalized and the nonlocal order parameters capable of distinguishing between the different phases of these theories are the Fredenhagen-Marcu operators~\cite{Fred_Mar_Quark,HuseFredMar}. We compute the latter operators in the different phases using time-independent perturbation theory. We show that the topological ordering exists in the Mott-insulator and charge-$2e$ superconductor phases of the system, while it disappears in the charge-$e$ superconductor phase.

Recent experimental endeavours in mesoscopic superconducting systems have shown a lot of promise towards detection of the MZM-s~\cite{Kouwen_experiment,Marcus_Majo_Exp,Nichele_Majo_2017,Lutchyn_Majo_exp_2017,Kouwen_Majo_2018, Wang_Majo_2018}.
Motivated by these developments, we also compute the topological gap using perturbation theory for small, but finite, $E_J$ and $E_M\ll E_C$. This  computation is relevant for near-term experimental implementation of the Majorana toric code since the experiments are most likely to be done for finite Josephson tunneling rate. 

The article is organized as follows. In section~\ref{sec:model}, we describe the microscopic Hamiltonian of a two-dimensional square lattice of MZM-s on mesoscopic islands and map it to a lattice gauge theory of $\mathrm{U}(1)$ matter and $\mathbb{Z}_2$ gauge fields. In Sec.~\ref{sec:toporder}, we introduce and compute the equal-time Fredenhagen-Marcu operator in the different phases of the system. In Sec.~\ref{sec:Heff}, we determine the influence of inter-island Cooper pair hopping on the toric code gap in the limit of dominant charging energy. In Sec.~\ref{sec:conclusion}, a concluding summary is provided. Details of calculations and a short review of lattice gauge order parameters are given in the appendices. 

\section{\label{sec:model}The model}
\begin{figure}
\subfigure[ ~$2$D lattice]{\includegraphics[width=0.35\textwidth]{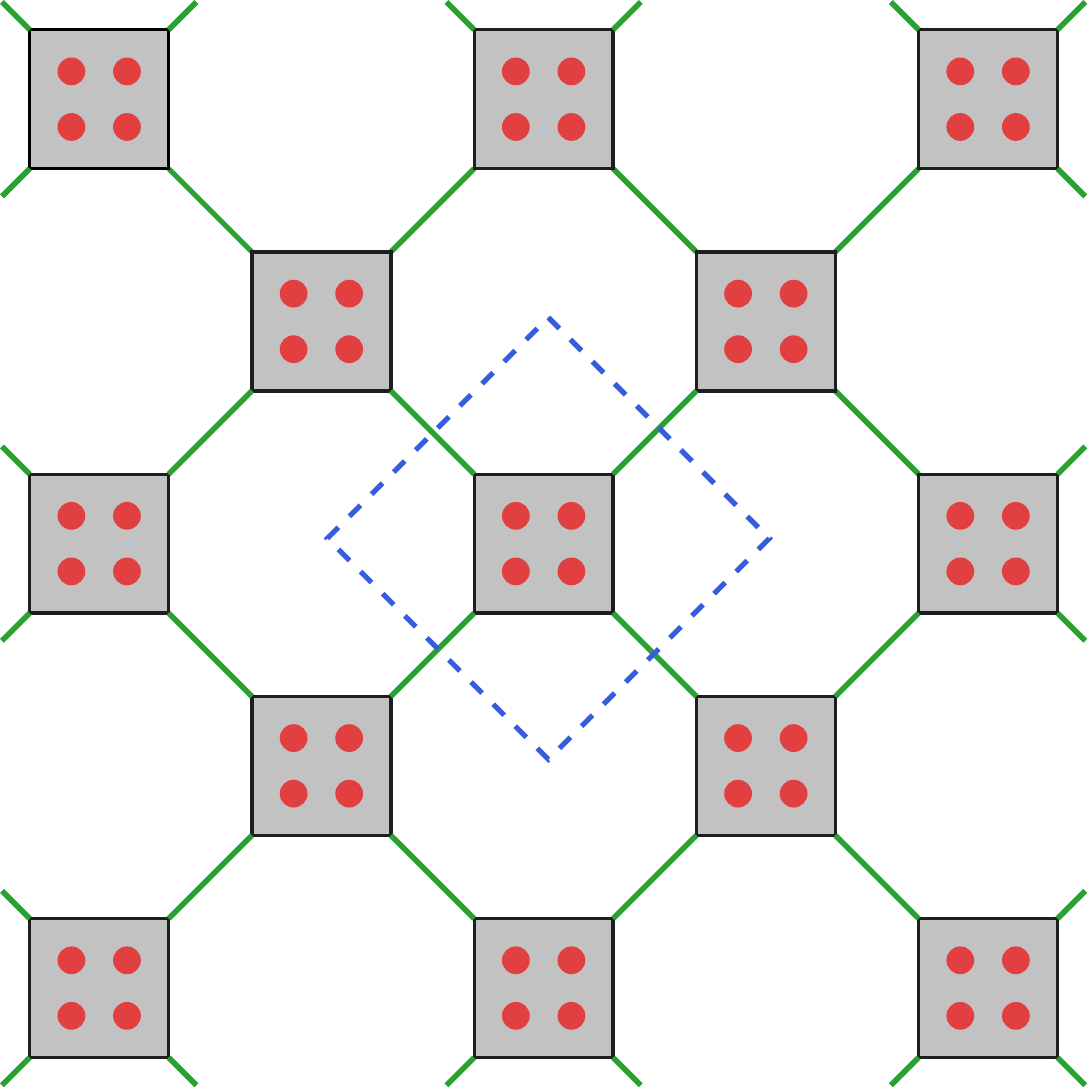}}
\subfigure[~ Unit cell]{\label{island}\includegraphics[width=0.3\textwidth]{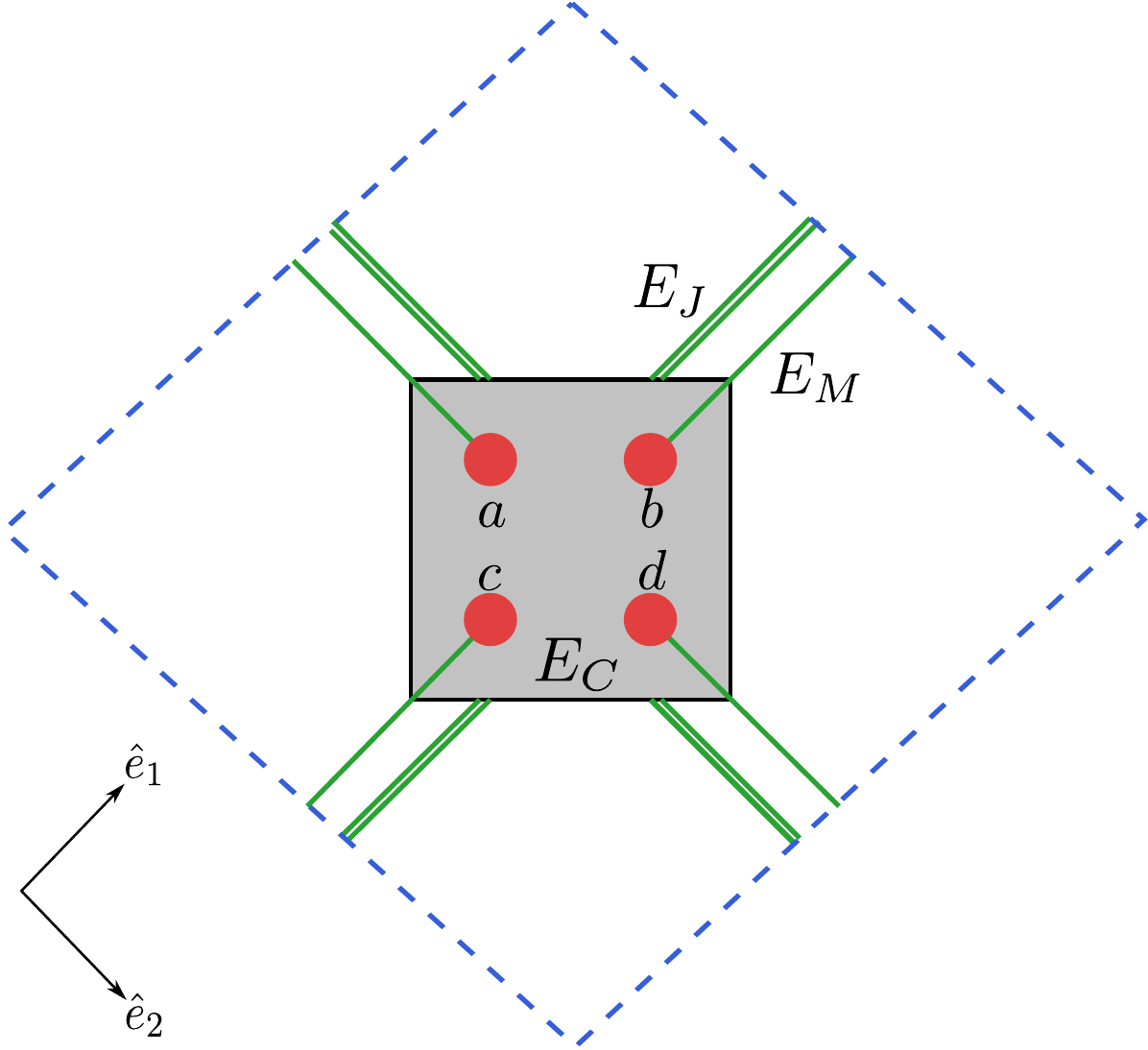}}%
\caption{Schematic of the Majorana toric code. Panel (a) shows the two dimensional square lattice of mesoscopic superconducting islands (gray) with nearest neighbour interaction (green). Panel (b) shows a unit cell of this lattice. On each island there are four Majorana zero modes (red). Each island has a finite charging energy ($E_C$). The nearest-neighbor interactions are due to Cooper pair tunnelling (at rate $E_J$) and Majorana-assisted single-electron tunnelling (at rate $E_M$).}
\label{microsketch}
\end{figure}
The microscopic model consists of a two dimensional lattice of mesoscopic superconducting islands, each carrying four MZM-s occuring as edge-modes of Kitaev wires~\cite{Kitaev_Chain}.  The nearest-neighbor islands are separated by tunnel junctions (see Fig.~\ref{microsketch}). The Hamiltonian of the system is given by $H = H_C + H_J + H_M$, where 
\begin{subequations}\label{microHam}
\begin{align}
	\label{hc}
	H_C &= 4 E_C \sum\limits_{j} n_j^2\, , \\\label{hj}
	H_J &= -E_J \sum\limits_{\langle j,k\rangle} \cos{\left( \phi_j-\phi_k \right)}\, ,\\\label{hm}
	H_M &=-E_M \sum\limits_{\langle j,k\rangle} i \gamma^j \gamma^k \cos{\left(\frac{\phi_j-\phi_k}{2}\right)}\, .
\end{align}
\end{subequations}
The first term, $H_C$, denotes the charging energy associated with each mesoscopic island. Here $E_C$ is the overall charging energy scale $e^2/2C$, where $C$ is the self-capacitance of each island. In contrast to the Cooper-pair box~\cite{KochTransmon, ChargeQubit},
here $n_i$ denotes excess number of fermions on each island and thus, can take both integer or half-integer values. 
The four MZM-s on each island are denoted by Hermitian operators $\gamma^j_\alpha$, $\alpha = a,b,c,d$, satisfying anticommutation relations:  $\{\gamma^j_\alpha,\gamma^k_\beta \} = 2\delta_{\alpha\beta}\delta_{jk}$. Even though they are not directly present in $H_C$, the MZM-s impose a gauge (parity) constraint satisfied by the physical state,
$\ket{\text{Phys}}$, of the system, given by~\cite{MajoTunnelFu,Roy2017,van_Heck_2012_constraint}
\begin{align}\label{microgauge}
	Q_j \ket{\text{Phys}} &= \ket{\text{Phys}} \, , \notag \\
	~\text{where} \quad &Q_j = -\gamma_{a}^j\gamma_{b}^j \gamma_{c}^j \gamma_{d}^j  ~\text{e}^{2 \pi i n_j} \, .
\end{align}
The second term, $H_J$, describes the coherent tunneling of Cooper pairs between neighboring islands, $E_J$ being the Josephson tunneling rate. Finally, the third term, $H_M$, describes the Majorana-assisted single-electron tunneling, $E_M$ being the relevant tunneling rate. The factor of $1/2$ in the argument of the cosine indicates that $1/2$ of the charge of the Cooper-pair is being transferred, while the fermionic operators keep track of the change in the fermion number parity~\cite{MajoTunnelFu, van_Heck_Hassler}. To avoid clutter, we have dropped the subscripts of the MZM operators in Eq.~\eqref{hm}.
Throughout this work, we consider the case of zero off-set charge. 

\begin{figure} 
\includegraphics[width=0.46\textwidth]{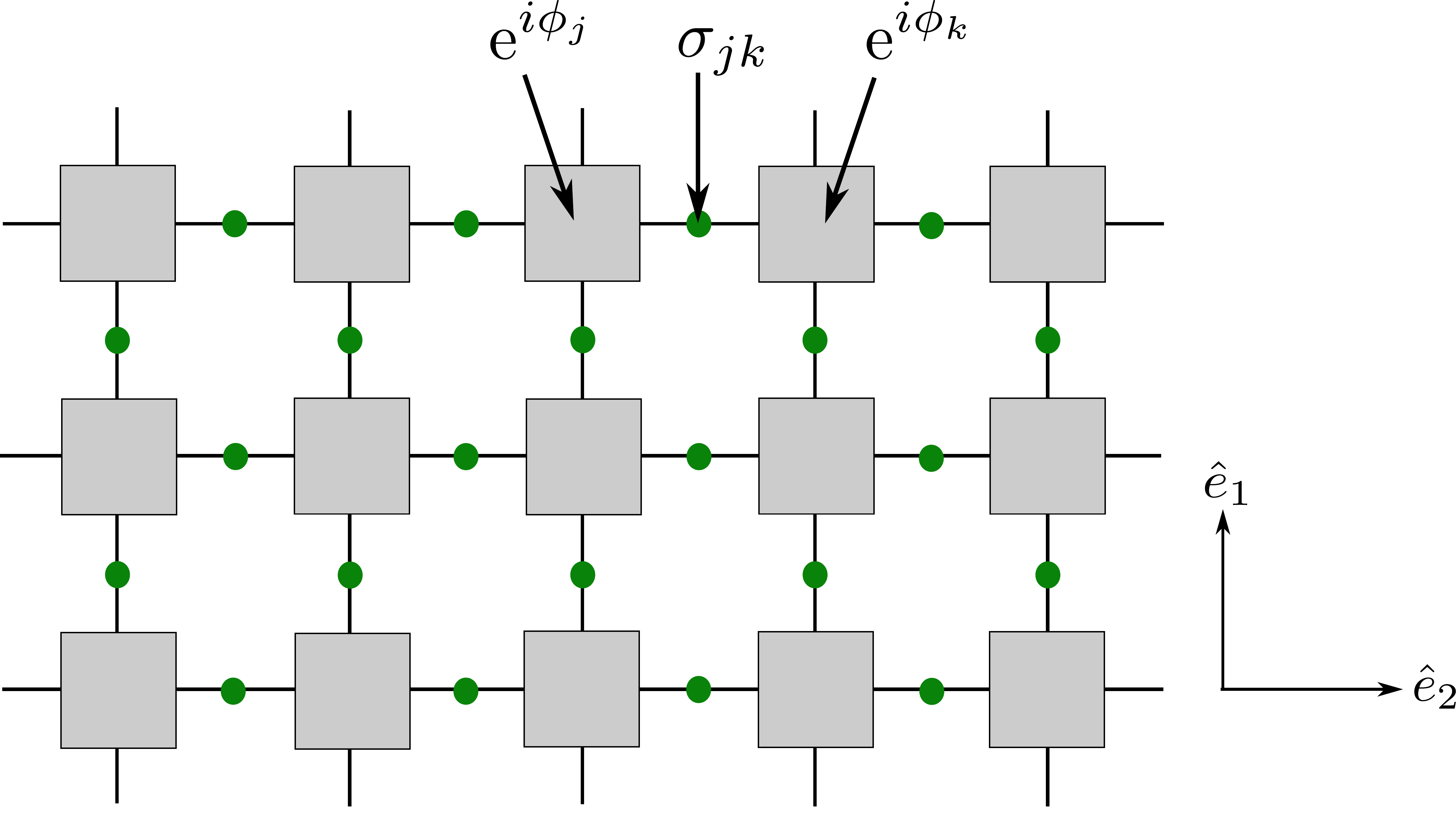}
\caption{Schematic of the two dimensional square lattice, with $\mathrm{U}(1)$ matter degrees of freedom $\text{e}^{i \phi_j}$ (gray squares) located at the nodes of the lattice and $\mathbb{Z}_2$ gauge degrees of freedom $\sigma_{jk}$ (green circles) placed on the links connecting nearest neighbouring nodes. This system Hamiltonian [Eq.~\eqref{HMTC}] realizes a $\mathbb{Z}_2$ lattice gauge theory with $\mathrm{U}(1)$ matter fields.}
\label{lattice_gauge_pic}
\end{figure}

In order to analyze the topological ordering in the system, we map the Hamiltonian in Eq.~\eqref{microHam} to that of interacting spins  coupled to quantum rotors. This is done by the transformation
\begin{equation}
	i \gamma^j \gamma^k ~\rightarrow~ \sigma_{jk}^z \quad \text{and} \quad -\gamma_{a}^j\gamma_{b}^j \gamma_{c}^j \gamma_{d}^j ~\rightarrow~ \prod_{+} \sigma^x \, , \label{Majotrafo}
\end{equation} 
using the bond algebraic approach~\citep{Cobanera_bondalgebra, Nussikov} (see Appendix~\ref{trafo} for details). This can be viewed as a Jordan-Wigner transformation followed by a duality transform and maps the product of the MZM-s located at the endpoints of a link to a spin placed centrally on it. This transformation keeps the number of degrees of freedom (DOF) invariant and the resulting Hamiltonian is given by (see Fig.~\ref{lattice_gauge_pic})
\begin{align} 
	H_{MTC} &= H_C + H_J + H_M \notag \\
	&= 4E_C \sum\limits_j n_j^2 - E_J\sum\limits_{\langle j,k\rangle}  \cos \left( \phi_{j}-\phi_{k} \right)  \notag \\\label{HMTC}
	&~~~~- E_M \sum\limits_{\langle j,k\rangle}\sigma^z_{jk} \cos \left( \frac{\phi_{j}-\phi_{k}}{2} \right).
\end{align}
The gauge constraint is transformed to
\begin{equation}\label{QMTC}
\quad Q_j = \text{e}^{2 \pi i n_j} \prod\limits_+ \sigma^x\, , 
\end{equation}
where in the last equation, $\prod_+\sigma^x$ indicates the product of the Pauli-X operators of the spins residing on the links incident at the $j^{th}$ island. 
Equation~\eqref{QMTC} is the discretized Gauss' law for the system~\cite{KogutIntro}, but unlike conventional electrodynamics, a $\mathrm{U}(1)$ matter field ($e^{i\phi_j}$), residing on the vertices, is coupled through its parity to a $\mathbb{Z}_2$ gauge field ($\sigma_{jk}^z$), residing on the links. The gauge fields do not have dynamics of their own since only $\sigma_{jk}^z$ operators occur in the Hamiltonian [Eq.~\eqref{HMTC}]. However, they acquire dynamics through the gauge constraint which couples them to the parity of the $\mathrm{U}(1)$ matter fields. 

The phase-diagram (see Fig.~\ref{phase_diag_rot}) of the system is rich and has been analyzed~\cite{Terhal2012, RoyTerhal2017, XuFu, Roy2017}. For small $E_J\ll E_M\ll E_C$, the system is in a toric code phase, while being a Mott insulator. Increasing $E_M/E_C$ keeping $E_J\ll E_M, E_C$, causes the system to undergo a 3D-XY quantum phase transition to a charge-$e$ superconductor phase. For $E_J\gg E_M\gg E_C$, the system is again in a toric code phase, while being a conventional superconductor. Increasing $E_M/E_C$, while keeping $E_J\gg E_M, E_C$, causes the system to undergo a 3D-Ising type quantum phase transition to a charge-$e$ superconductor state. These different phase-transitions are connected by a couple of tricritical points and first-order transitions. Finally, a 3D-XY type quantum phase transition separates the Mott-insulator and the conventional superconductor phases. 

While charge signatures of the different phases have been analyzed in the earlier works, the existence of the toric code has been indirectly inferred. In the next section, we compute non-local order parameters to provide direct evidence of the presence/absence of toric code ordering. 

\section{\label{sec:toporder}Analysis of topological ordering using The Fredenhagen-Marcu operator}
For models with global symmetry, the phases are distinguished by a {\it local} order parameter (the celebrated symmetry-breaking paradigm of Landau). In contrast, for models with gauge symmetry, which cannot be broken~\cite{Elitzur_1975}, the different phases can be differentiated by {\it non-local} order parameters~\cite{NUSSINOV_non_local}. For a pure gauge theory, like the Ising gauge theory, the Wilson loops have their characteristic area (perimeter) law decay in the confined (deconfined) phases (see Ref.~\onlinecite{KogutIntro} and references therein). However, the presence of dynamical matter fields, alongside gauge fields, introduces additional complications. The matter fields potentially screen the fluctuations of the gauge fields and can cause the Wilson loop of the gauge fields to decay with a perimeter law in both the confined and the deconfined phases (throughout we refer to confinemenet/deconfinement of the external charges of the gauge field) of the system, {\it e.g.}, the Ising gauge theory in the presence of Ising matter fields~\cite{FradShenk}. The Majorana toric code, as was shown in the previous section, can be viewed as such a theory, where dynamical $\mathrm{U}(1)$ matter fields interact with $\mathbb{Z}_2$ gauge fields.

For these theories, the notion of the Wilson loop has to be generalized to distinguish between the different phases of the model and the relevant non-local order parameter is the Fredenhagen-Marcu (FM) operator~\cite{Fred_Mar_Quark}, proposed initially for models of particle physics and has been used for condensed matter systems~\cite{HuseFredMar}. 
In contrast to the Wilson loop which has the same behavior in both confined and deconfined phases, the FM operator has the desired feature of an order parameter: it is zero (nonzero) in the confined (deconfined) phases of the system. There are three different operator formulations depending on the space-time orientation of the non-local operators~\cite{HuseFredMar} and we use the equal-time formulation (a short review of the Wilson loop and different FM order parameters is given in Appendix~\ref{app:LGOP}).
\begin{figure}
\begin{center}
\subfigure[~Equal-time version]{\includegraphics[width=0.29\textwidth]{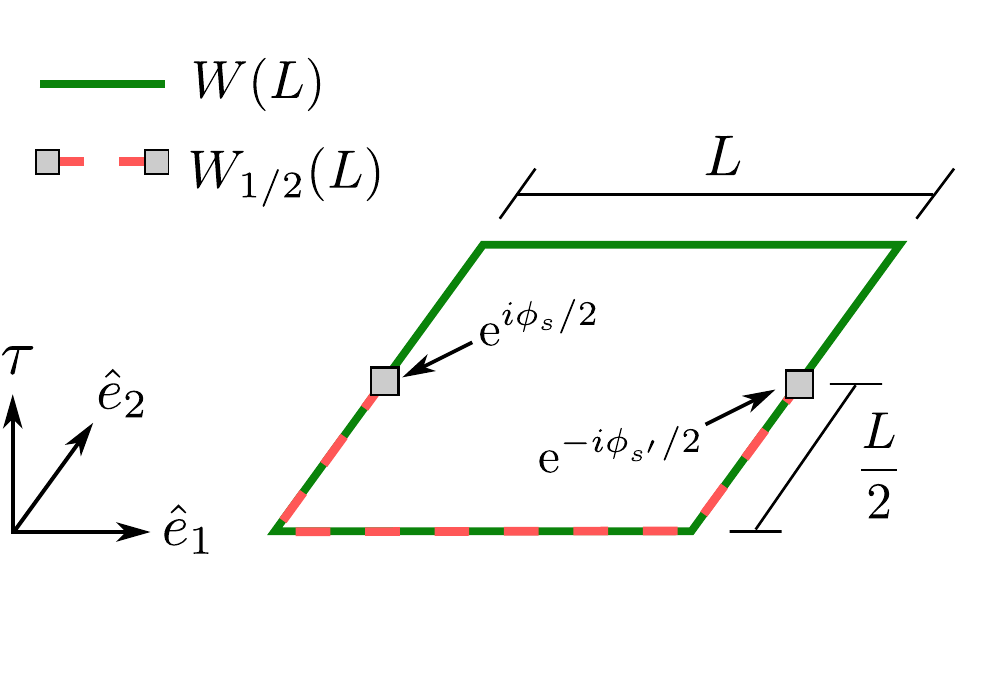}\label{fig:Fred_Mar_a}} \hfill
\subfigure[~Fredenhagen-Marcu]{\includegraphics[width=0.17\textwidth]{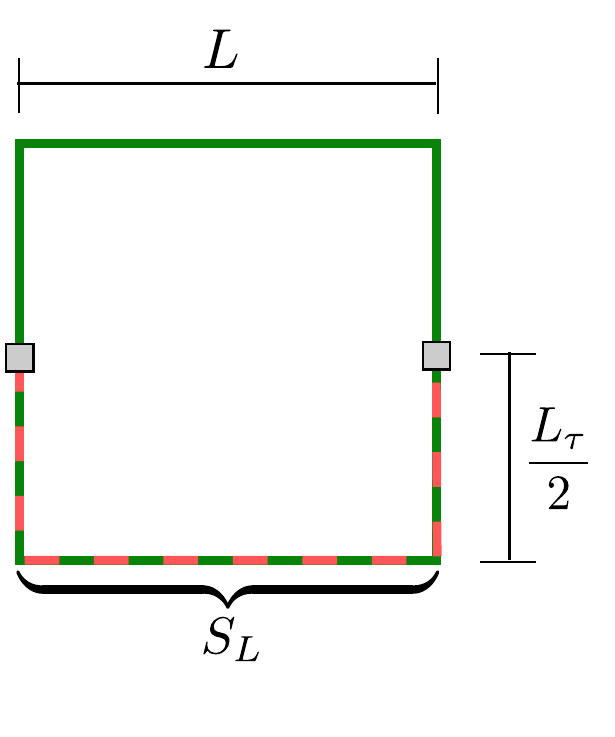} \label{fig:Fred_Mar_b}}
\caption{The Fredenhagen-Marcu order parameter. Panel (a) shows the equal-time version. The green line denotes the gauge spins $\sigma_{l}^z$ along a square of side length $L$, a space-space Wilson loop. The red dashed line depicts a half Wilson loop, with matter degrees of freedom (gray) at the endpoints. Panel (b) shows the operator originally proposed by Fredenhagen and Marcu, where one of the directions is the imaginary time direction $\tau$. The strings of gauge spins at a given imaginary time are denoted as $S_L$ [see Appendix~\ref{app:LGOP}, Eq.~\eqref{space_time_Woperator}].}
\label{fig:Fred_Mar_sketch}
\end{center}
\end{figure}
The Wilson loop for the system is given by 
\begin{align}\label{Wilson_for_FM}
	W(L)&= \bra{G}\prod_{l \in \mathcal{C}} \sigma_l^z\ket{G}\, ,
\end{align}
where $\sigma^z_l$ are the Pauli-Z operators of the spins residing on the contour $\mathcal{C}$, the latter being a square of side length $L$ [cf. Fig.~\ref{fig:Fred_Mar_a}], and $|G\rangle$ is (one of) the ground state(s) of the system. In order to construct the FM operator, we need also the modified half Wilson loop. It comprises a product of a string of gauge field operators along the contour $\mathcal{C}_{1/2}$, terminated by two matter field operators at sites $s$ and $s'$, given by
\begin{equation}\label{Wilson_half_FM}
	W_{1/2}(L) =  \bra{G}e^{i\phi_{s}/2} \prod_{l \in \mathcal{C}_{1/2}} \sigma_l^z ~e^{-i\phi_{s'}/2}\ket{G}\, .
\end{equation} 
The equal-time FM operator, $R(L)$, is given by 
\begin{align}\label{FM_equal_time}
	R(L) = \frac{W_{1/2}(L)}{\sqrt{W(L)}}\, .
\end{align}
In the limit of an infinitely large loop, one can show that (see Appendix ~\ref{app:LGOP} for details)
\begin{equation}\label{condecon}
	\lim\limits_{L \to \infty} R(L) = \begin{cases} 0\, , & \mathbb{Z}_2\ \text{deconfined,}\\
	{\rm const.}, & \mathbb{Z}_2\ \text{confined.} \end{cases} \,
\end{equation}
The construction of the FM operator can be understood as follows. In order to correctly diagnose the signatures of the confining/deconfining phases, a `correct string tension', which determines the potential between the gauge charges, needs to be computed. To that end, consider half a Wilson loop [see Fig.~\ref{fig:Fred_Mar_a}]. Just by itself, this quantity is not gauge-invariant, but can be made so by gluing matter fields at the end-points. In the confining phase, this half-Wilson loop with matter fields at the ends, $W_{1/2}(L)$, acquires a finite
expectation value, whereas in the deconfining phase, it vanishes; see Appendix B. The division by a square-root of the Wilson loop is done to distill the dependence of the diagnostic on the coupling constant from the size of the loop.

Intuitively, both the Wilson loop and the FM operator can be best interpreted in the space-time formulation [see Fig.~\ref{fig:Fred_Mar_b}]. In this formulation, loops extend in one spatial and one temporal direction. In absence of matter charges, the Wilson loop is a quantity purely involving gauge DOF and is related to the potential between external, static, charges of the gauge fields. In the presence of dynamical matter charges which can potentially screen the interaction between the gauge field charges, the Wilson loop is no longer capable of correctly diagnosing the potential between the gauge charges. This is remedied by the FM operator which is defined in terms of the  dynamical matter charges to compensate for the screening effect. More on the FM operator can be found in Appendix~\ref{app:LGOP} and Ref. \onlinecite{HuseFredMar}.

The Majorana toric code describes interacting $\mathrm{U}(1)$ matter fields and $\mathbb{Z}_2$ gauge fields in two spatial and one (Euclidean) time direction. The toric code phase corresponds to the system being in the $\mathbb{Z}_2$ deconfined phase. Below, we compute $R(L)$ in the different phases of the Majorana toric code using time-independent perturbation theory. Since we evaluate the FM operator expectation value at equal times, it is sufficient to consider the expectation value in the ground state of the system. 
\begin{figure}
\begin{center}
\includegraphics[width=0.47\textwidth]{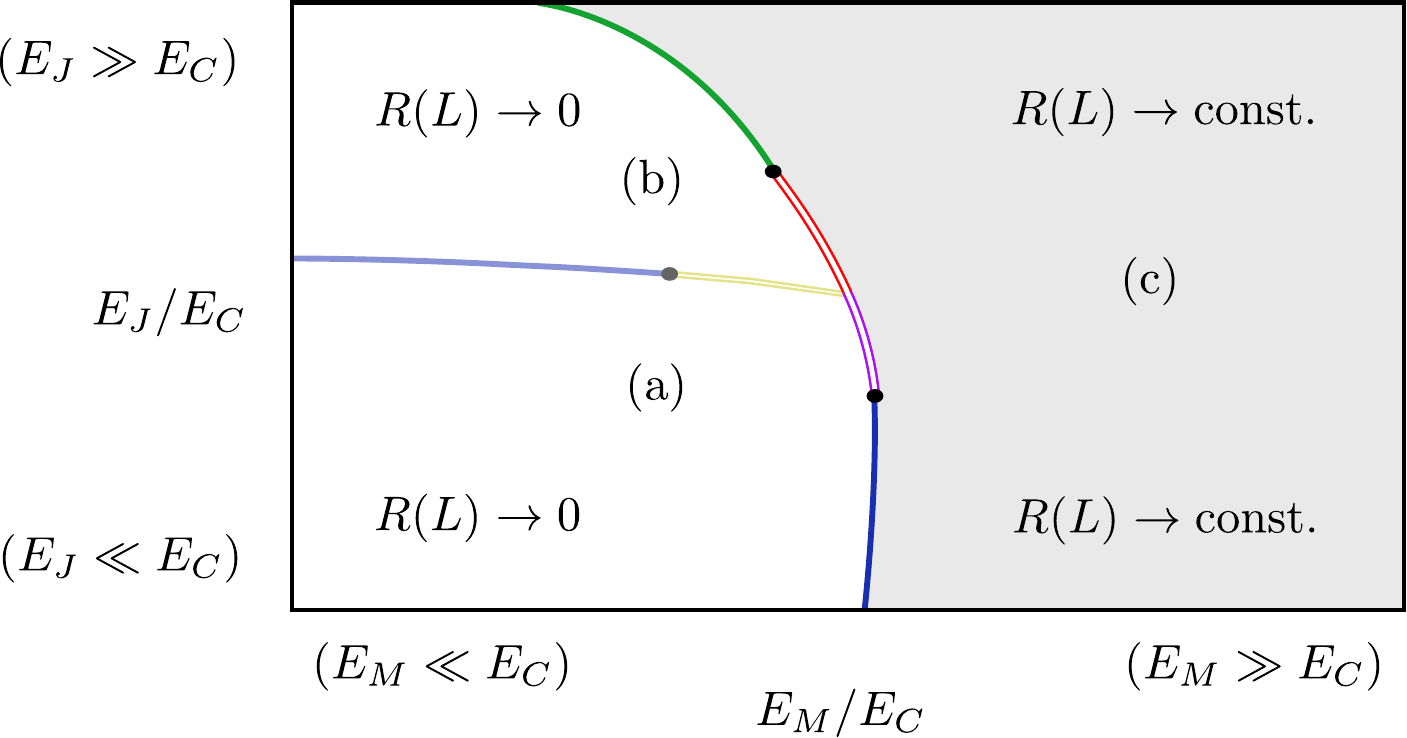}
\caption{Phase diagram of the Majorana toric code. For $E_J\ll E_M\ll E_C$, the system is in a Mott-insulator phase [denoted by (a)]. The Fredenhagen-Marcu (FM) operator is: $R(L)\rightarrow0$, indicating a $\mathbb{Z}_2$ deconfined (toric code) phase. Increase of $E_J/E_C$ causes the system to undergo a pure matter (Higgs) phase transition of 3D-XY type to a charge-2$e$ superconductor [denoted by (b)], while remaining in the toric code phase with $R(L)\rightarrow0$. Increase of $E_M/E_C$ from either (a) or (b) causes the toric code ordering to disappear $R(L)\rightarrow \text{const.}$ and the system transforms into a charge-$e$ superconductor [denoted by (c)]. The nature of the phase-transition is either 3D-XY [(a) $\rightarrow$ (c)] or 3D-Ising [(b) $\rightarrow$ (c)]. The 3D-Ising and 3D-XY lines terminate at tricritical points, before turning first order. The boundary of the toric code phase is inferred from the field theory~\cite{RoyTerhal2017, Roy2017} and the perturbation theory calculation of the next section. }
\label{phase_diag_rot}
\end{center}
\end{figure}

Before providing the details of the computation, we provide a summary of the findings in Fig.~\ref{phase_diag_rot}. The Mott-insulator [phase (a)] and the charge-2$e$ superconductor [phase (b)] are in the toric code phase, indicated by $R(L)\rightarrow0$, while the charge-$e$ superconductor [phase (c)], with $R(L)\rightarrow$ constant, does not exhibit toric code ordering. This result also indicates that there are no additional phase-transitions as $E_J$ is varied between $0$ and $\infty$ for large $E_M$. Thus, the results of our computations of the FM operator support the earlier field theory computations.\\  
\\

\subsection{\label{sec:strongcharge}Dominant charging energy}
\begin{figure}
\begin{center}
\includegraphics[width=0.47\textwidth]{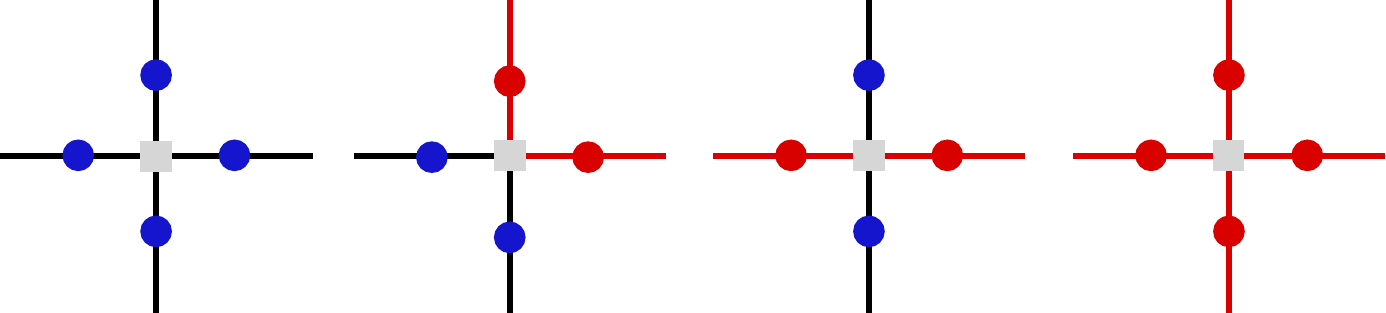}
\caption[Allowed node configurations]{Allowed node configurations that satisfy $\prod_+ \sigma^x=1$. The fixed matter fields (gray squares) are depicted for completeness, but are irrelevant for the topological ordering computations. We focus on a specific node $i$, where the adjacent gauge degrees of freedom (circles) are displayed in the $x$-basis. The colors blue/red indicate the gauge field $\sigma_{ij}$ to be in the $+1/-1$ eigenstate of $\sigma^x_{ij}$. To be valid, the configuration needs to contain an even number of $-1$ eigenstates. To obtain the complete set, the four (two) possible rotations of the second (third) graph have to be considered as well. Coloring the respective links in red, it is evident that the strings of spins in $-1$ eigenstate have to form closed loops.}
\label{loop_flow_pic}
\end{center}
\end{figure}

First, we consider the parameter regime where the Josephson as well as the single-electron tunneling rate is small compared to the charging energy: $E_J,E_M\ll E_C$. It is evident that the charge (excess number of fermions) on each island is a good quantum number and the system is a Mott insulator~\cite{RoyTerhal2017}. The unperturbed Hamiltonian is $H_C$, given in Eq.~\eqref{HMTC}, with the unperturbed ground state being $|G\rangle = |n_i=0\rangle$, $\forall i$. As a consequence, the gauge constraint reduces to 
\begin{equation}\label{chargeless_gauge}
	\qquad\prod\limits_{+}\sigma^x \ket{G}=\ket{G}\, .
\end{equation}
Imposing this constraint allows eight configurations of gauge fields at each node, as shown in Fig.~\ref{loop_flow_pic}, where out of the four spins incident at each vertex, an even number of them are in $\sigma^x = +1$ (blue circles) and $\sigma^x=-1$ (red circles). 
If we imagine coloring the links of the lattice with red whenever the spins are in the $\sigma^x=-1$ eigenstate, then all possible closed red loop configurations of the lattice are allowed and the ground state is a loop condensate~\cite{ToricCode_anyons, fradkin_2013}. The degeneracy of the ground state is four-fold, due to the presence of the four non-contractible loops on a torus. 
While calculating using perturbation theory~\footnote{For the perturbation theory calculation, it is sufficient to consider one of the four degenerate ground states since their splitting is exponentially suppressed in the system size.}, we impose the constraint in Eq.~\eqref{chargeless_gauge} directly on the unperturbed ground state and obtain equal weight ($w$) superpositions of all possible closed loop configurations $\mathcal{S}$ 
\begin{equation}\label{GS_EC}
	\ket{G} = w\sum\limits_{s \in \mathcal{S}}~ \lvert n_i=0,~ s \rangle = w\sum\limits_{s \in \mathcal{S}}\ket{s}\, .
\end{equation}
Including the single and double charge hopping from Eq.~\eqref{HMTC} as perturbations $V= H_M+H_J$, time independent calculations yield 
\begin{align}
	R(L) = \text{e}^{-\ln \left( 4\frac{E_C}{E_M}\right) L} ~ \overset{L \rightarrow \infty}{\longrightarrow} ~ 0\, ,
\end{align} 
indicating the presence of a $\mathbb{Z}_2$ deconfined (toric code) phase for dominant charging energy (see Appendix~\ref{app:FM_calc_1} for details of the computation). 

\subsection{\label{sec:largeJoseph}Dominant Josephson tunneling rate}
Next, we analyze the case when the Josephson tunneling rate is the strongest: $E_J \gg E_M, E_C$, when the system can be mapped to an effective Ising gauge theory, as explained below. In this limit, the phase differences on the  islands are pinned to multiples of $2 \pi$ up to a gauge choice. In this reduced subspace, the $\mathrm{U}(1)$ DOF behaves like an effective $\mathbb{Z}_2$ DOF. The operator $\text{e}^{\pm i \phi_i /2}$ acts like an effective Pauli-Z operator:  $\tau_i^z$ whose eigenvalues can be $\pm1$ depending on whether the $\mathrm{U}(1)$ DOF are pinned to even/odd multiples of $2\pi$. The charging energy term induces quantum fluctuations of the phase, leading to constant energy shifts and $2 \pi$-phase slips on the nodes. Thus, the exponent of the number operator $n_i$ acts like a Pauli-X operator: $\text{e}^{2 \pi i n_i} ~\rightarrow ~ \tau^x_i$, where  $\comm{\tau_i^z}{\tau_j^x}=\comm{\text{e}^{\pm i \phi_i /2}}{\text{e}^{2 \pi i n_j}}$.  Disregarding the constant contribution, the charging energy term can effectively be incorporated into the $\mathbb{Z}_2$ formulation as $H_{C,i}~\rightarrow~ -\Delta \tau_i^x$ leading to the  effective Hamiltonian
\begin{figure}
\begin{center}
\includegraphics[width=0.5\textwidth]{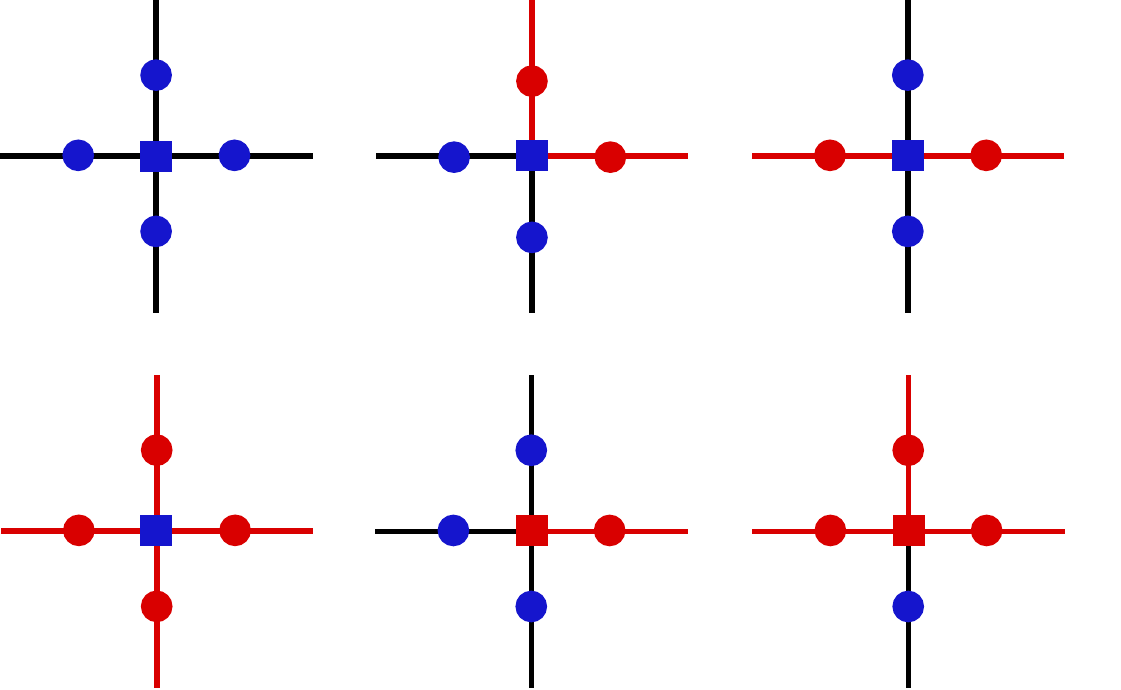}
\caption[Allowed node configurations in the presence of matter.]{Allowed node configurations that satisfy $\tau_i^x\prod_+ \sigma^x=1$. We focus on a specific node $i$, where the adjacent gauge degrees of freedom (circles) as well as the matter degrees of freedom (squares) are displayed in the $x$-basis. Note that we consider the matter after the effective mapping on to the $\mathbb{Z}_2$ variables. The colors blue/red indicate the gauge and matter $\sigma_{ij}$ and $\tau_i$ to be in the $+1/-1$ eigenstate of $\sigma^x_{ij}$ and $\tau^x_i$. To be valid, the configuration needs to contain an even number of $-1$ eigenstates. To obtain the complete set, the four (two, four, four) possible rotations of the second (third, fifth, sixth) graph have to be considered as well. By coloring the respective links also in red, it is evident that the $-1$ eigenstates form open loops terminated by flipped matter fields.}
\label{loop_flow_pic_matter}
\end{center}
\end{figure}
\begin{align}\label{HZ2}
	H &=  -\Delta \sum\limits_i \tau^x_i - E_M \sum\limits_{<i,j>}\tau^z_i \sigma^z_{ij } \tau^z_{j } \\
	\text{and} \qquad Q_i &= \tau_i^x \prod\limits_{+} \sigma^x\, .\label{gauge_for_Ising}
\end{align} To determine the dominant coupling dependence of $\Delta$, assume the slips to occur separately on each island. This approximation is analogous to the one dimensional case treated in Ref.~\onlinecite{vanHeck1DMTC}. The coupling of independent slips is then calculated as the tunnelling amplitude in the cosine potential using semi-classical WKB approximation~\cite{Landau_Lifschitz, Terhal2012}, leading to  $\ln{\Delta} \propto -\sqrt{E_J / E_C}$. The obtained Hamiltonian is that of the 2D quantum $\mathbb{Z}_2$ gauge theory. This is seen by choosing the London gauge, where the matter fields are unity: $\tau^z_i=1$ and using the gauge constraint to replace $\tau^x_i = \prod_+ \sigma^x$. Switching to the dual lattice and rotating the basis yields
\begin{align}
\label{hz3}
	H = -\Delta \sum\limits_i \prod\limits_{\Box} \sigma^z - E_M \sum\limits_{<i,j>}\sigma^x_{ij}\, .
\end{align}
This directly verifies the limit considered in Ref.~\onlinecite{Terhal2012}, where the toric code phase was predicted to exist for $\Delta\gg E_M$. Increasing $E_M/\Delta$ caused the system to undergo a 3D-Ising type phase-transition to a non-toric code phase. The argument of Ref.~\onlinecite{Terhal2012} is based on working in a dual picture and analyzing the phase-transition of the quantum Ising model. Below, we provide a more direct proof of the toric code ordering by computing the expectation value of the FM operator in the ground state of the gauge-invariant Hamiltonian given in Eq.~\eqref{HZ2}. 

\subsubsection{Large charging energy: $E_C \gg E_M$}
Since in this limit, $\Delta\gg E_M$, from Eq.~\eqref{HZ2}, the matter fields are all pinned: $\tau^x_i=1$. From Eq.~\eqref{gauge_for_Ising}, only those configurations of the gauge field are allowed which satisfy the gauge constraint. Those are closed loops where loops are formed by the gauge field in $\sigma^x = -1$ state, exactly as for the case $E_C\gg E_M, E_J$ analyzed in Sec.~\ref{sec:strongcharge}. Thus, the ground state is given by
\begin{equation}
\ket{G} = w \sum\limits_{s \in \mathcal{S}}~ \lvert \tau^x_i=1,~ s \rangle = w\sum\limits_{s \in \mathcal{S}}\ket{s},
\end{equation}
which is again a four-fold degenerate loop-condensate. Time independent perturbation theory then yields
\begin{align}
	R(L) = \text{e}^{-\ln \left( 4\frac{\Delta}{E_M}\right) L} ~ \overset{L \rightarrow \infty}{\longrightarrow} ~ 0 \, ,
\end{align}

again indicating a toric code phase. 

\subsubsection{Large single-electron tunneling rate: $E_M \gg E_C$} 
In this limit, $E_M\gg\Delta$ and from the gauge constraint in Eq.~\eqref{gauge_for_Ising}, the valid vertex configurations displayed in Fig.~\ref{loop_flow_pic_matter} allow for open (red) loops of $\sigma^x=-1$ terminated by flipped parity ($\tau_i^x=-1$). The unperturbed ground state is
\begin{equation}\label{Gopen}
	\ket{G} = w\sum_{m \in \mathcal{M}} \ket{m} \, ,
\end{equation}
where $\mathcal{M}$ denotes the set of all open loop configurations. Note that closed loops are treated as open loops with coinciding start- and endpoint such that those are also contained in $\mathcal{M}$. In contrast to the closed loop condensate, where only an even number of non-contractible loops are generated, Eq.~\eqref{Gopen} contains also odd numbers of non-contractible loops in the equal weight superposition leading to a non-degenerate ground state.  Computing the expectation value of the FM operator in the ground state, we find (see Appendix~\ref{app:FM_calc_2} for details) \begin{align}
	R(L) = \text{e}^{- \frac{1}{16}{(\frac{\Delta}{E_M})}^2 } ~ \overset{L \rightarrow \infty}{\longrightarrow} ~ \text{const.} \, 
\end{align}
This indicates that the system is in a $\mathbb{Z}_2$ confined phase [see Eq.~\eqref{condecon}] and thus, does not have the toric code ordering.

\subsection{\label{sec:weakJoseph}Dominant single electron tunneling rate}
We consider the limit of $E_M \gg E_C, E_J$ and we restrict ourselves to the case of $E_J=0$ since the limit of large $E_J$ was already analyzed in the previous section. For $E_M \rightarrow \infty$ in Eq.~\eqref{HMTC}, we find that $\phi_i-\phi_j = 0$ or $2\pi$ depending on whether $\sigma_{ij}^z = 1$ or $-1$. Assuming a fixed configuration $\lbrace \sigma_{ij}^z \rbrace$, the phases are again pinned to $\phi_i = 2\pi m_i$, where $m\in\mathbb{Z}$ . Thus, we can apply the same mapping as in the previous seciton to map the $\mathrm{U}(1)$ DOF to the Ising variable $\tau_i^z$. Furthermore, the charging term induces quantum fluctuations such that $H_{C,i} \rightarrow \Delta ' \tau_i^x$. Assuming independent slips, $\Delta'^2$ can be estimated by the tunnelling event in the $4\pi$ periodic cosine potential using the WKB method~\cite{Landau_Lifschitz}. This yields $\ln{\Delta'} \propto \-\sqrt{E_M / E_C}$.
Thus, the Hamiltonian takes the same form as in Eq.~\eqref{HZ2} only differing by an adjusted tunnelling rate $\Delta '$. It has the same non-degenerate unperturbed ground state resulting in the Fredenhagen-Marcu operator to be evaluated as
\begin{equation}
R(L) = \text{e}^{- \frac{1}{16}{(\frac{{\Delta '}}{E_M})}^2 } ~ \overset{L \rightarrow \infty}{\longrightarrow} ~ \text{const.} \, 
\end{equation}
Thus, the system is again in a $\mathbb{Z}_2$ confined phase and there is no toric code ordering. The summary of the our findings is given in Fig. \ref{phase_diag_rot}.\\

\section{\label{sec:Heff}Stability of the Toric Code in the dominant charging energy regime}
\begin{figure}
\begin{center}
\subfigure[]{\label{fig:pert_diag_a}\includegraphics[width=0.176\textwidth]{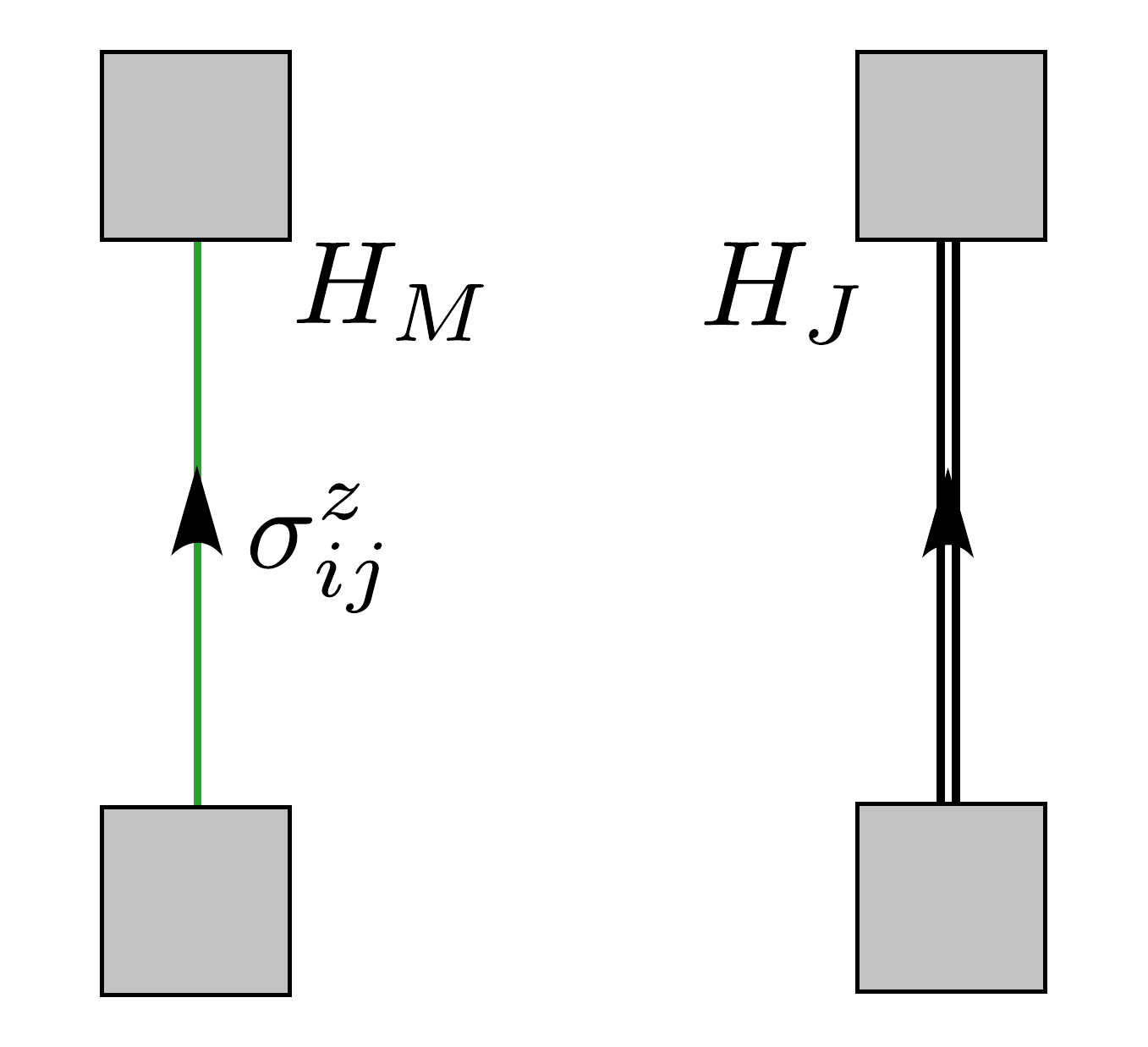}} \hfill
\subfigure[]{\label{fig:pert_diag_b}\includegraphics[width=0.176\textwidth]{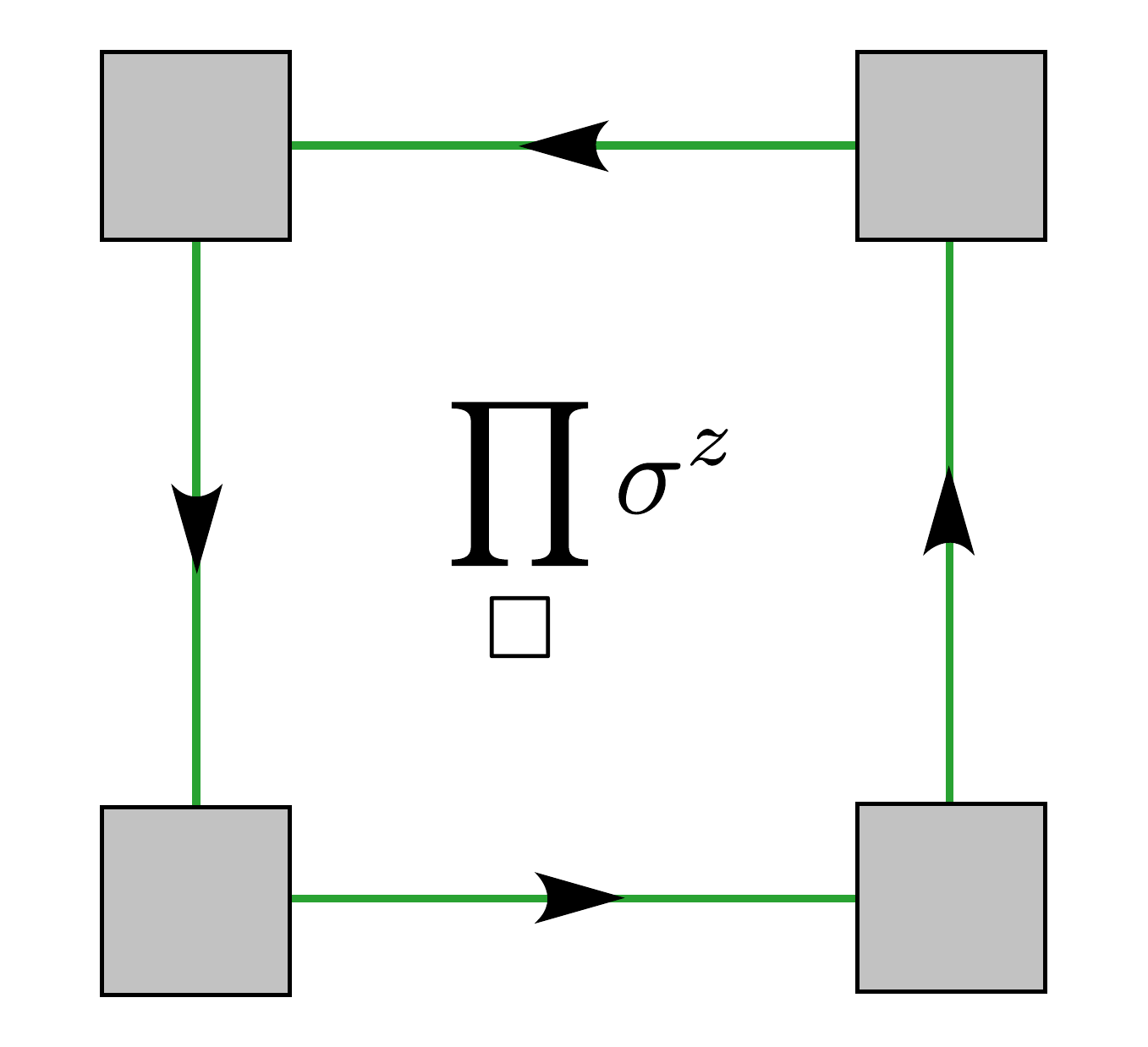}}\\
\subfigure[]{\label{fig:pert_diag_c}\includegraphics[width=0.176\textwidth]{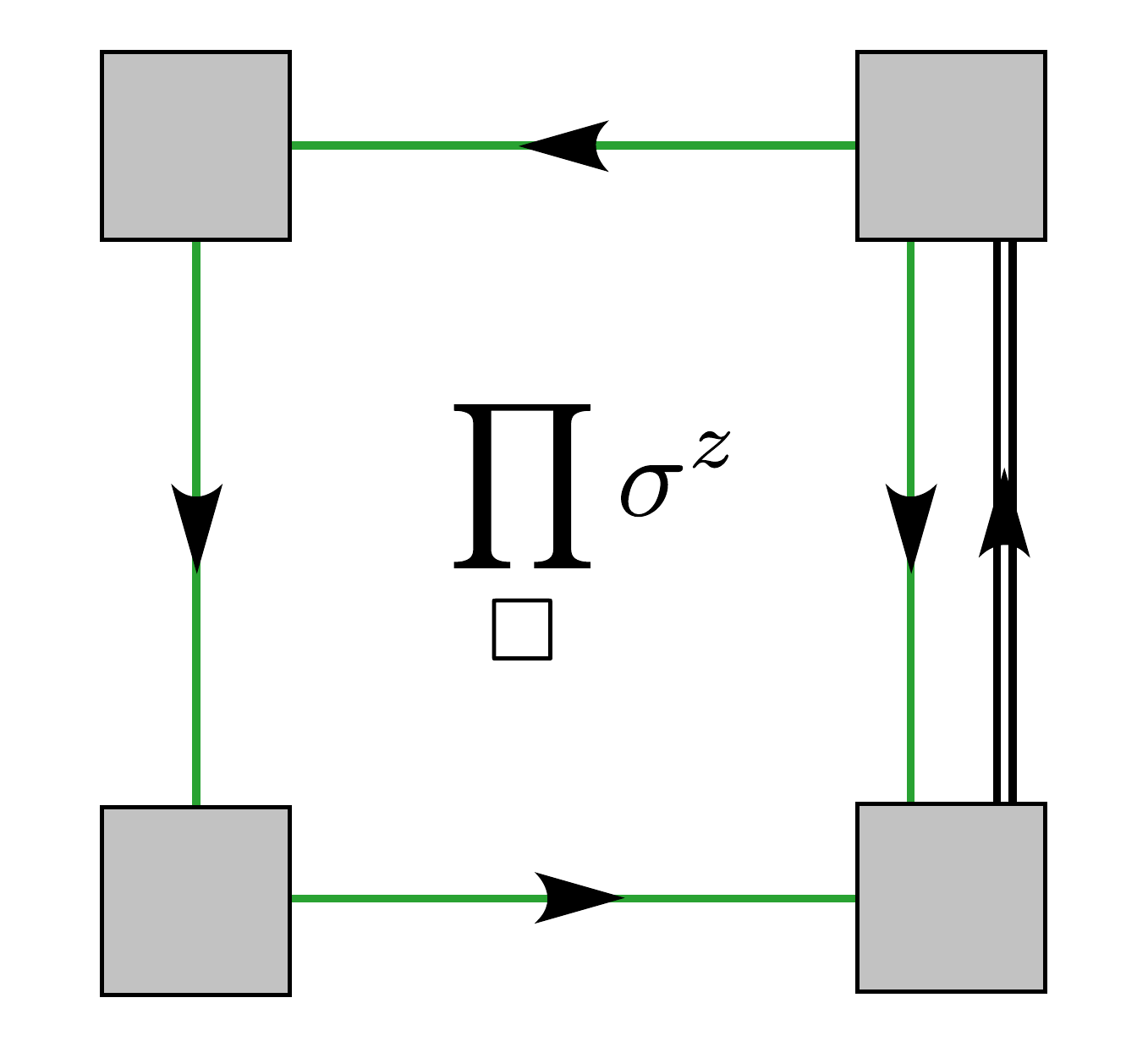}}\hfill
\subfigure[]{\label{fig:pert_diag_d}\includegraphics[width=0.29\textwidth]{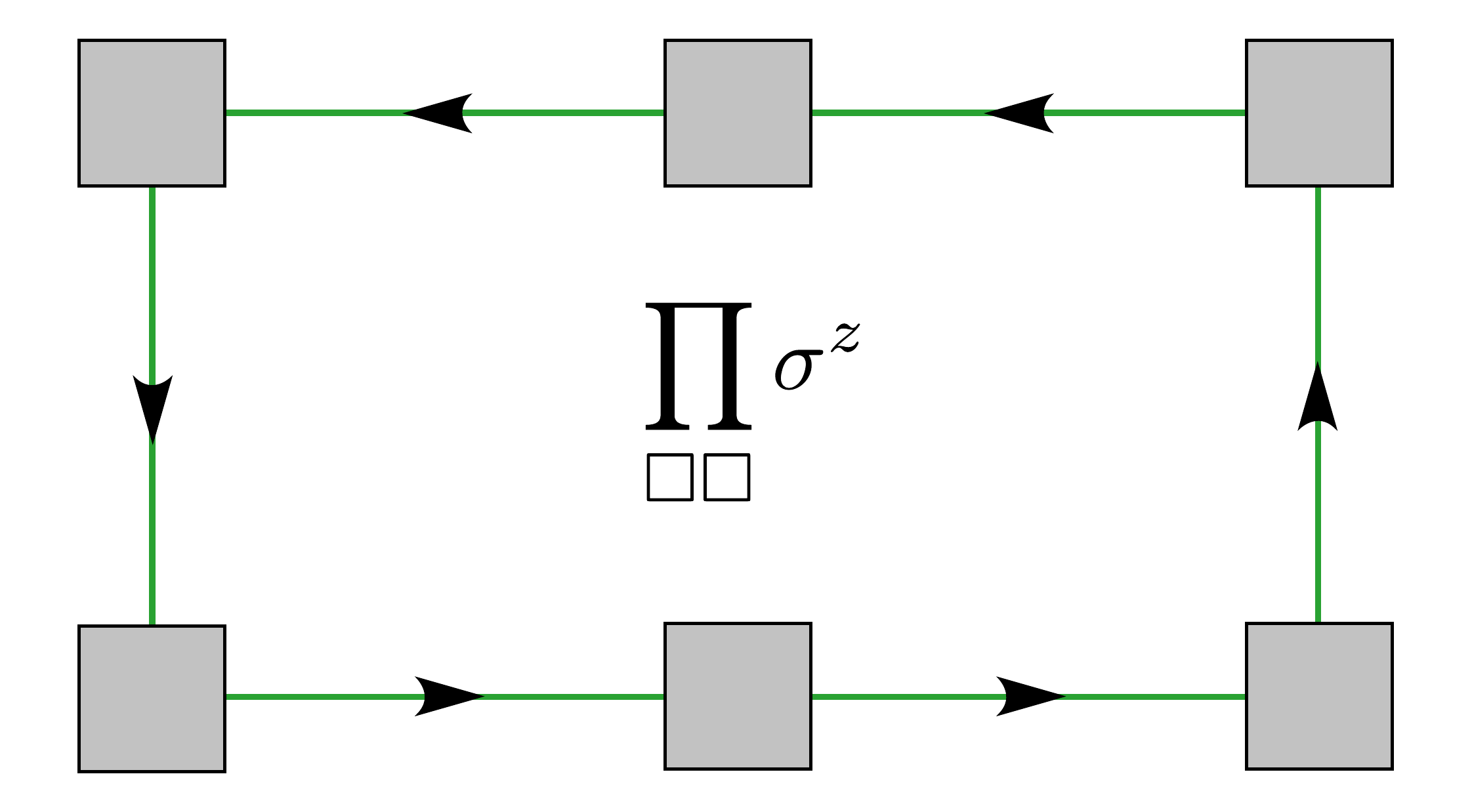}}
\caption{Spatial visualization of single perturbations (a), where the line indicates a link of the direct lattice. On the left is the tunnelling of a single electron from node $j$ to $i$, which lowers the charge on island $j$ by $1/2$ (in numbers of Cooper pairs) and adds this charge to node $i$. Additionally, the gauge field on the link between those nodes flips (colored green). On the right, a Cooper pair tunnels between the nodes. This involves a charge transfer of one. The diagrams (b) through (d) show non-constant fourth to sixth order diagrams contributing to the effective Hamiltonian.}
\label{fig:pert_diag}
\end{center}
\end{figure}

In this section, we compute the topological gap of the toric code phase perturbatively in $E_J/E_C$ and $E_M/E_C$.
We consider the Hamiltonian [Eq.~\eqref{HMTC}] of the system in the Mott insulator phase (a), where $E_C \gg E_J,E_M$ and determine the effective Hamiltonian up to sixth order in perturbation theory, for details see Appendix~\ref{sec:perturbdescr}. The goal of this computation is to provide a quantitative prediction of the effect of Cooper pair tunneling on the toric code gap. The different virtual processes that contribute to the effective Hamiltonian are depicted in Fig.~\ref{fig:pert_diag_a}. The number of lines indicates the number of single charges that are transferred and the color green denotes the appearance of $\sigma^z$ for the link. Every valid contribution to $H_{\text{eff}}$ is given by a process that starts and ends in a charge less state [see the projectors $P_-$ in Eq.~\eqref{Heffsum}]. 

The lowest order contribution to the effective Hamiltonian that is not a constant, is obtained at fourth order in the perturbation theory. This process is depicted in Fig.~\ref{fig:pert_diag_b}. This is due to the process which transfers a single charge around the smallest loop on the lattice, a plaquette. The $\mathbb{Z}_2$ (gauge) DOF are carried around by the $\mathrm{U}(1)$ charge transfer operators $e^{\pm i\phi/2}$ and lead to a non-constant contribution to the Hamiltonian [see Fig.~\ref{fig:pert_diag_b}]. The respective fourth order term in operator formulation is given by~\cite{LandauPlugge}
\begin{equation}\label{H4plaq}
	H_{\text{eff}}^{(4)}~ =~ \text{const.}- \frac{5}{16}\frac{E_M^4}{E_C^3} \sum\limits_{i} \prod\limits_{\Box} \sigma^z \, .
\end{equation}
In combination with the gauge constraint [Eq.~\eqref{QMTC}], this gives the toric code Hamiltonian.\\

To determine the leading order contribution in $E_J$, we have to go to fifth order in perturbation and find the process shown in Fig.~\ref{fig:pert_diag_c}. 
A single charge effectively flows around a plaquette. However, in contrast to Fig.~\ref{fig:pert_diag_b}, a single-electron tunnels opposite to the direction of the overall charge-flow around the plaquette. This opposite movement of the single electron is compensated by a transfer of Cooper pair along the direction of the overall charge-flow around the plaquette. Since the gauge operators appear independent of the flow direction [cf. Eq.~\eqref{HMTC}] it yields a plaquette term. The contribution to the effective Hamiltonian is given by
\begin{align}\label{Heff_five}
	H_{\text{eff}}^{(5)} = \text{const.}-\frac{61}{144} \frac{E_M^4 E_J}{E_C^4}~ \sum\limits_i \prod\limits_{\Box} \sigma^z \, .
\end{align}
Thus, for small $E_J$, the hopping of Cooper pairs increases the gap of the toric code and stabilizes the toric code space. In contrast to the Cooper pair box Hamiltonian $H_C+H_J$, where the sign of $E_J$ leaves the spectrum invariant, it becomes important due to the introduction of $H_M$.
Instead, Eq.~\eqref{HMTC} is only left invariant by the transformation
\begin{equation}
	E_{M} \rightarrow -E_{M} \quad \text{and} \quad \phi_k \rightarrow \phi_{k} + 2\pi\, , \quad \forall~ k \in K \, ,
\end{equation}
where $K$ is a set of indices containing next nearest neighbouring islands, one sub lattice in a bipartite lattice.
The phase diagram (cf. Fig.~\ref{phase_diag_rot}) is thus not expected to be symmetric along the $E_J=0$ axis, while the transition lines approach the $E_M=0$ at a right angle.\\  

At sixth order, the non-constant contributions can be separated in two terms:
\begin{equation}
	H_{\text{eff}}^{(6)} = \text{const.} + H_{\text{eff}}^{(6,1)} +H_{\text{eff}}^{(6,2)}\, .
\end{equation}
First, there are next to leading order plaquette terms, which we compute to be 
\begin{equation}
	H_{\text{eff}}^{(6,1)} = -\left[ \frac{259}{1728}\frac{E_M^6}{E_C^5} + \frac{24509861}{71124480} \frac{E_M^4 E_J^2}{E_C^5} \right] \sum\limits_i \prod\limits_{\Box} \sigma^z\, .
\end{equation}
Second, we find an additional process depicted in Fig.~\ref{fig:pert_diag_d}. A single charge can flow around the perimeter of two adjacent plaquettes, which is the second smallest cycle in the square lattice. The gauge fields on the perimeter get thereby flipped. The sum of the relevant diagrams for this contribution is calculated to be
\begin{equation}
	H_{\text{eff}}^{(6,2)} = -\frac{63}{256}\frac{E_M^6}{E_C^5} \sum\limits_i ~\biggl( ~\prod\limits_{\Box\Box} \sigma^z + \prod\limits_{\substack{ \text{$\Box$} \\ \text{$\Box$}}} \sigma^z ~\biggr)\, .
\end{equation}
The $\sigma^z$ on the shared link can be added for clarity as they square to identity. This term introduces a nearest neighbour interaction between adjacent plaquettes. The interaction commutes with the single plaquette term and results in an additional stabilization of the ground state space~\cite{Terhal2012}. \\

\section{\label{sec:conclusion}Conclusion}
To summarize, we have analyzed the signatures of topological ordering in the different phases of the Majorana toric code. To that end, we mapped the system onto a model of interacting $\mathrm{U}(1)$ matter fields and $\mathbb{Z}_2$ gauge fields. We computed a non-local order parameter, the equal-time Fredenhagen-Marcu order parameter to capture the topological ordering in the system. Our calculations confirm that the Mott-insulator and the charge-2$e$ superconductor phases of the system show an emergent toric code, while the charge-$e$ superconductor phase does not exhibit toric code ordering. Our results are compatible with the earlier field theory predictions in the different parts of the phase-diagram. Furthermore, we computed the topological gap of the toric code in the presence of finite Cooper pair tunneling and showed  that a small amount of Cooper pair tunneling actually stabilizes the toric code phase since the topological gap increases with a small amount of Cooper pair tunneling. 

\section{\label{sec:ack}Acknowledgements}
A.Z. and F.H. acknowledge funding by the Deutsche Forschungsgemeinschaft (DFG, German Research Foundation) under Germany's Excellence Strategy -- Cluster of Excellence Matter and Light for Quantum Computing (ML4Q) EXC 2004/1 -- 390534769. A.R. acknowledges the support of the Alexander von Humboldt foundation. 
\appendix

\section{\label{trafo} From Majoranas to spins}

In this section, we construct the duality transformation given in Eq.~\eqref{Majotrafo} based on the bond algebraic approach from Refs.~\onlinecite{Nussikov, Cobanera_bondalgebra}. First, we define $\mathcal{L}$ the set of all nearest neighbour bonds, $\mathcal{I}$ the set of all nodes and the operator $P_i = -\gamma_{a}^i\gamma_{b}^i \gamma_{c}^i \gamma_{d}^i$ to compactify the notation. 
The set of bonds generating the bond algebra is then given by
\begin{align}
	i \gamma^k \gamma^l\, , \quad \forall (k,l) \in \mathcal{L} \qquad \text{and} \qquad P_i\, , \quad \forall i \in \mathcal{I}\, .
\end{align}

Making use of the Clifford algebra one can show that the bond algebra $\mathcal{A}$ contains the intensive relations, \textit{i.e}., independent of the lattice size,
\begin{itemize}
	\item[I)] $(i \gamma^k \gamma^l)^2 = P_j^2 = \boldsymbol{1}\, , \qquad \forall (k,l) \in \mathcal{L}\, , \forall j \in \mathcal{I}\, , $
	\item[II)] $\acomm{P_k}{i \gamma^k \gamma^l} = \acomm{P_l}{i \gamma^k \gamma^l} = 0\, , \qquad \forall (k,l) \in \mathcal{L}\, ,$
	\item[III)] all remaining bond combinations commute.
\end{itemize}
For periodic boundary conditions, there remains one extensive, \textit{i.e.}, lattice size dependent, relation
\begin{itemize}
	\item[IV)] $\prod\limits_{j \in \mathcal{I}} P_j = \alpha \prod\limits_{(k,l) \in \mathcal{L}} i \gamma^k \gamma^l$\, ,
\end{itemize}
where $\alpha=\pm 1$ accounts for the different orderings of the Majorana operators on both sides of the equation. This last equation considers the global $\mathbb{Z}_2$ symmetry of the Majoranas. We want to reproduce the same bond algebra $\mathcal{A}$ with alternative bonds, where there is a spin-$1/2$ degree of freedom placed on the links between two islands (cf.~Fig.~\ref{lattice_gauge_pic}). This DOF is represented by the regular Pauli operators $\sigma_{ij}^z,\sigma_{ij}^y,\sigma_{ij}^x$ obeying the Pauli algebra. The sizes of the Hilbert spaces agree as there are two Majorana modes per island which can be occupied or not in case of the original bonds and two links per island containing a spin-$1/2$ DOF which can be up or down in the alternative bond description. The new set of bonds can be defined as
\begin{align}
	&\sigma_{kl}^z\, , \qquad \forall (k,l) \in \mathcal{L} \notag \\ 
	\text{and} \qquad &\tilde{P}_i = \prod\limits_{+_i}\sigma^x \equiv \prod\limits_{j:(i,j) \in \mathcal{L}} \sigma_{ij}^x\, , \quad \forall i \in \mathcal{I}\, ,
\end{align}
so that $\tilde{P}_i$ is given by the product of $\sigma^x$ of all four links emanating from island $i$. With the Pauli algebra it can be verified that the alternative bonds satisfy the intensive relations
\begin{itemize}
	\item[I)] $(\sigma_{kl}^z)^2 = \tilde{P}_j^2 = \boldsymbol{1}\, , \qquad \forall (k,l) \in \mathcal{L}\, , \forall j \in \mathcal{I}\, , $
	\item[II)] $\acomm{\tilde{P}_k}{\sigma_{kl}^z} = \acomm{\tilde{P}_l}{\sigma_{kl}^z} = 0\, , \qquad \forall (k,l) \in \mathcal{L}\, ,$
	\item[III)] all remaining bond combinations commute.
\end{itemize}
In order to fulfill the extensive relation, we choose one arbitrary, but fixed island and set $\tilde{P}_{0} = \prod\nolimits_{+_{0}}\sigma^x \cdot ( \alpha \prod_{(k,l) \in \mathcal{L}} \sigma_{kl}^z ) $. With this choice, the intensive relations still hold, as the additional prefactor commutes with all remaining bonds and we obtain
\begin{itemize}
	\item[IV)] $\prod\limits_{j \in \mathcal{I}} \tilde{P}_j = \alpha \prod\limits_{(k,l) \in \mathcal{L}} \sigma_{kl}^z\, ,$
\end{itemize}
since all $\sigma^{x}$ square to identity. Now that the Hilbert space sizes are equal and both the intensive as well as the extensive relations are satisfied, we can write down the new model description as in Eq.~\eqref{HMTC}. Both Hamiltonians are linked by a unitary transformation, see Ref.~\onlinecite{Nussikov}.

\section{Lattice gauge order parameters\label{app:LGOP}}
In this section, we give a short review on the Wilson loop as well as an in-depth discussion of the Fredenhagen-Marcu operator to make the article self-contained.

\subsection{The Wilson loop operator}
\begin{figure}[tbp] 
\begin{center}
\subfigure[~Space-space Wilson loop]{\includegraphics[width=0.29\textwidth]{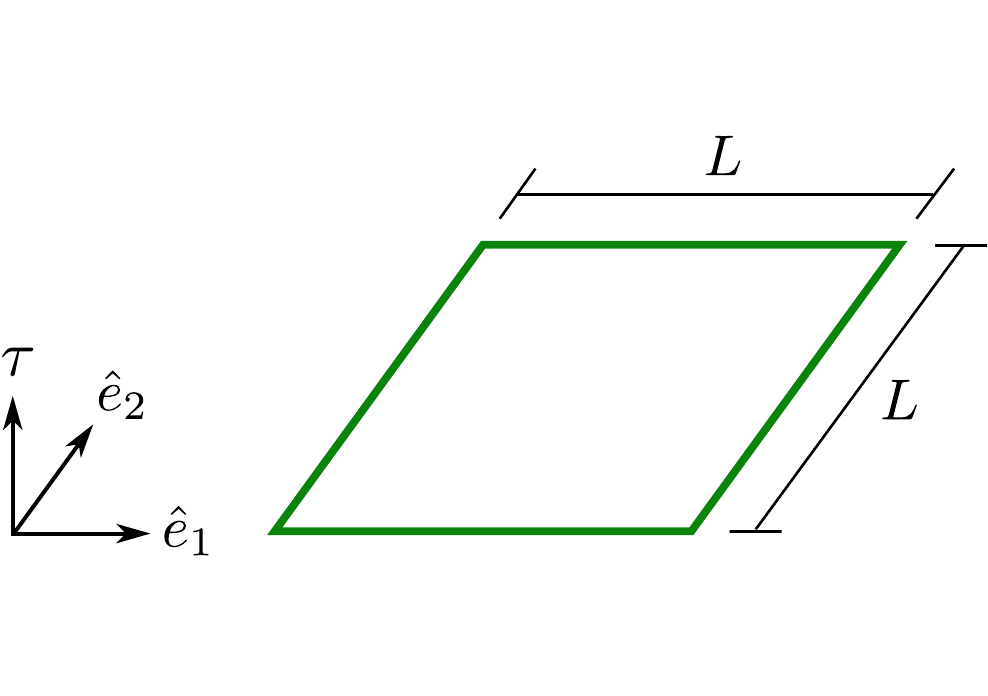}\label{fig:Wilson_a}}\hfill
\subfigure[~Space-time Wilson loop]{\includegraphics[width=0.16\textwidth]{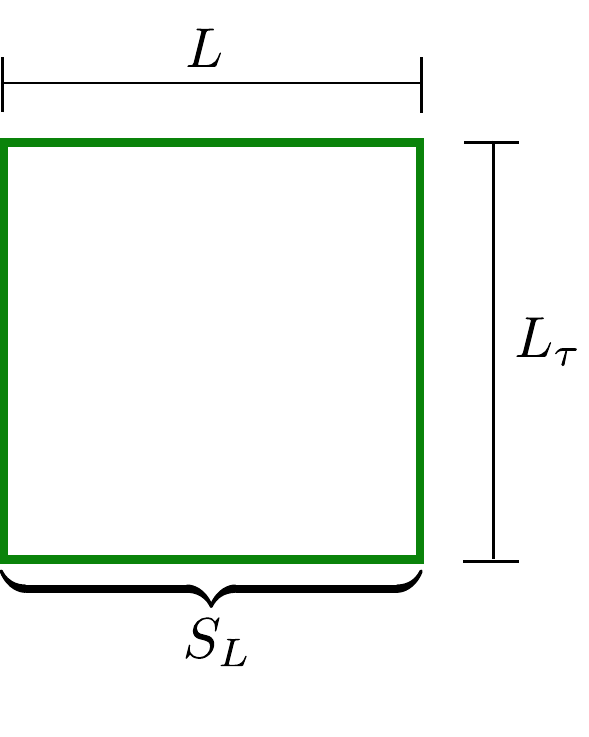} \label{fig:Wilson_b}}
\caption{Visualization of the Wilson loop order parameter. In panel (a) there is the space-space Wilson loop. The green line denotes the gauge spins $\sigma_{l}^z$ along a square of side length $L$. In panel (b) the space-time Wilson loop is depicted, where one of the directions is the imaginary time direction $\tau$. A different length $L_{\tau}$ in temporal direction allows for the anisotropic limit. The strings of gauge spins at a given imaginary time are denoted as $S_L$.}
\label{fig:Wilsons}
\end{center}
\end{figure}

The Wilson loop was introduced by Wegner~\cite{Wegner1971} for the isotropic Ising gauge model, where it distinguishes the area from perimeter law~\cite{KogutIntro}. For anisotropic models, \textit{e.g.} with continuous (imaginary) time direction as in Eq.~\eqref{HMTC}, distinct operator formulations for the Wilson loop are possible (cf. Fig.~\ref{fig:Wilsons}). 

The space-space operator formulation of the Wilson loop is given by
\begin{align}\label{Wilson_intro}
	W(L)&= \bra{G}\prod\limits_{l \in \mathcal{C}} \sigma_l^z\ket{G}\, .
\end{align}
To discuss the contour $\mathcal{C}$, we have to look at Figure~\ref{fig:Wilson_a}. The contour is a square of side length $L$, thus the $L$ dependence on the left hand side of Eq.~\eqref{Wilson_intro} stems from this implicit dependence in $\mathcal{C}$. The space-space formulation yields the ground state expectation of a square loop of side length $L$. It is a gauge invariant measure of the correlation between gauge fields, that are a distance $L$ apart.\\

To construct the space-time formulation, we define the imaginary time dependent gauge spin operators per link $l$ as~\cite{HuseFredMar}
\begin{equation}\label{im_time_propa}
	\sigma^z_{l}(\tau) = \text{e}^{-H \tau} ~\sigma^z_{l} ~\text{e}^{H \tau} \, .
\end{equation}
Furthermore, we apply the temporal gauge $\sigma_{l'}^z = 1$ for all links $l'$ in the imaginary time direction. The length of the loop in imaginary time is given by $L_{\tau} = v \mathcal{T}$, where we assume the velocity $v$ to be unity. This results in a space-time operator formulation of
\begin{align}\label{space_time_Woperator}
	W(L,\mathcal{T})&= \bra{G}S(L,0)~~S(L,-\mathcal{T}) \ket{G}\, .
\end{align}
The strings are defined as $S(L, \tau) = \prod\nolimits_{l \in S_L} \sigma_l^z(\tau)$, where $S_L$ is the curve of links depicted in Fig.~\ref{fig:Wilson_b}. Those are the remaining spatial lines of the loop at different time slices. For convenience, we choose the strings to lie at times $\tau =-\mathcal{T}$ and $\tau =0$. This operator is connected to the potential $V$ between two external static, infinitely heavy, charges of the gauge field placed a distance $L$ apart~\cite{KogutIntro} with
\begin{align}
	V(L) \propto - \lim\limits_{\mathcal{T}  \rightarrow \infty} \frac{1}{\mathcal{T}} \ln \left[ W(L, \mathcal{T}) \right] \, .
\end{align} 
The latter relation leads to the labelling of phases in the Ising gauge model as
\begin{equation}
	W(L) = \begin{cases}\text{e}^{-\alpha A}\, , & \mathbb{Z}_2 \text{ confined},\\
	\text{e}^{-\beta P}\, , & \mathbb{Z}_2 \text{ deconfined,} \end{cases} 
\end{equation} 
where $A$ $(P)$ denotes the area (perimeter) of the Wilson loop and $\alpha, \beta$ are model dependent prefactors. In the presence of dynamical matter this diagnostic breaks down, if the matter fluctuations introduce screening and thus perimeter law everywhere~\cite{FradShenk}. \\

\subsection{The Fredenhagen-Marcu operator}

The interpretation of the equal-time Fredenhagen-Marcu operator defined in Eqs.~\eqref{Wilson_for_FM}--\eqref{FM_equal_time} is analyzed using the lattice gauge theory of Ising gauge coupled to Ising matter, where we follow the treatment in Ref.~\onlinecite{HuseFredMar}. We consider a system of spins $\tau / \sigma$ on the nodes/links of a two dimensional square lattice. The respective Hamiltonian is given by
\begin{align}
	 H &= -\Delta \sum\limits_i \prod\limits_{\Box} \sigma^z -\frac{1}{\Delta}\sum_{<i,j>} \sigma_{ij}^x  \notag \\
	 &\quad~  - \lambda \sum\limits_{<i,j>} \tau^z_i \sigma_{ij}^z \tau^z_j -  \frac{1}{\lambda} \sum\limits_{i} \tau_i^x \, .\label{HIsing_gauge_matter}
\end{align}
	
\begin{figure}[tbp] 
\begin{center}
\subfigure[ ]{\includegraphics[width=0.42\textwidth]{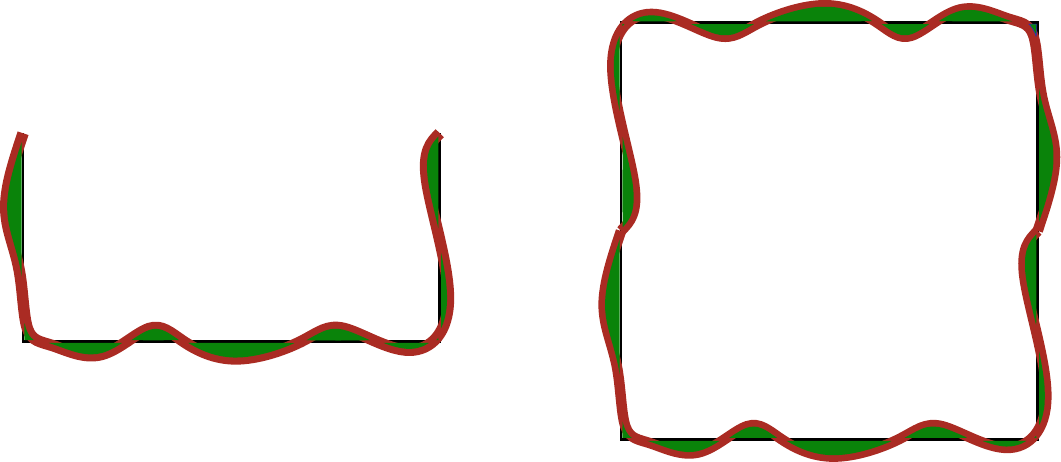}\label{fig:surface_a}} \hspace{20pt}
\subfigure[ ]{\includegraphics[width=0.42\textwidth]{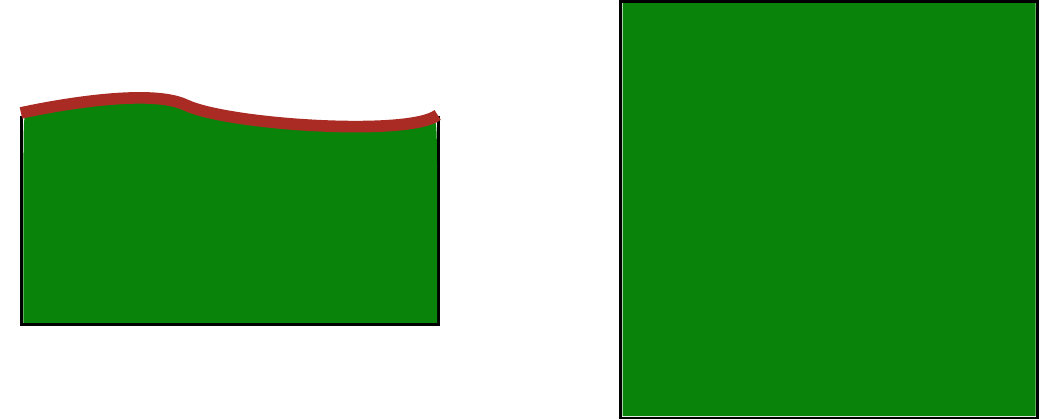} \label{fig:surface_b}}
\caption[Graphical evaluation of the Fredenhagen-Marcu operator]{Graphical evaluation of the equal-time version of the Fredenhagen-Marcu order parameter. In panel (a) open and closed loop are covered mostly with edges (red) as opposed to surfaces (green). In panel (b) the closed loop is covered completely with surfaces, while for the open loop edges have to be introduced that close the loop. The figure is inspired from Ref.~\onlinecite{HuseFredMar}.}
\label{fig:surface_edge}
\end{center}
\end{figure}

We refer to nearest neighbour hopping of matter DOF [proportional to $\lambda$ in Eq.~\eqref{HIsing_gauge_matter}] as edges and aligned gauge DOF around plaquettes [proportional to $\Delta$ in Eq.~\eqref{HIsing_gauge_matter}] as surfaces. Since the perimeter of a square of side length $L$ is considerably smaller than its area for large $L$, the system will generate the (half) Wilson loop both by mainly constructing its perimeter with edges.  More generally, we say that large loops will always be dominantly covered by edges as opposed to surfaces, as long as $\lambda \gtrapprox \Delta$ [cf. Fig.~\ref{fig:surface_a}]. Therefore, the numerator and denominator in Eq.~\eqref{FM_equal_time} have the same scaling behaviour, except when surfaces are much cheaper than edges. This is the case, when the matter field is considerably heavy, while the gauge field is ordered $\Delta \gg \lambda$. Then the closed loop gets filled with surfaces. The open loop is equivalently covered with surfaces, but the remaining ($L$ dependent) line has to be covered by expensive edges [cf. Fig.~\ref{fig:surface_b}]. Thus, the numerator decays much faster than the denominator. Central to these scaling behaviours is the necessity of matter flow (edges) in the generation of the half Wilson loop. In conclusion, we have that
\begin{equation}
	\lim\limits_{L \to \infty} R(L) = \begin{cases} 0\, , & \Delta \gg \lambda\, ,\\
	\text{const.,} & \text{else.} \end{cases} 
\end{equation}
Either the numerator decays faster or the scaling is equal up to a constant.
By taking the $\lambda \rightarrow 0$ limit, we can rewrite Eq.~\eqref{HIsing_gauge_matter} as
\begin{align}
	H &= -\Delta \sum\limits_i \prod\limits_{\Box} \sigma^z -\frac{1}{\Delta}\sum_{<i,j>} \sigma_{ij}^x    -  \frac{1}{\lambda} \sum\limits_{i} \prod\limits_+ \sigma^x\, ,
\end{align}
where we replaced $\tau^x_i = \prod\nolimits_+ \sigma^x$ via the gauge constraint. If also $\Delta \rightarrow \infty$, we find an emergent toric code. Using this we can infer
\begin{equation}
	\lim\limits_{L \to \infty} R(L) = \begin{cases} 0\, , & \mathbb{Z}_2\text{ deconfined,}\\
	\text{const.}, & \mathbb{Z}_2\text{ confined.} \end{cases} 
\end{equation}
This connection can be made more explicit by considering the original Fredenhagen-Marcu order parameter, where one spatial direction is exchanged for the imaginary time direction [cf.~Fig.~\ref{fig:Fred_Mar_b}]. 

We construct the operator formulation along the lines of Ref.~\onlinecite{HuseFredMar}. We assume that the two matter spins $\tau_s$ and $\tau_{s'}$ lie at the same time slice, which we set to $\tau=0$. Choosing the temporal gauge and performing the imaginary time propagation analogously to Eq.~\eqref{im_time_propa}, we find that
\begin{equation}
	W_{1/2}(L,\mathcal{T}) =  \bra{G}\tau^z_{s}(0)\tau^z_{s'}(0)~ S(L,-\mathcal{T}/2) \ket{G}\, .
\end{equation}
The gauge flux generated from a charge at $s$ and anti-charge at $s'$ gets introduced at time $-\mathcal{T}/2$ and the respective charges are created at time zero. Using a unit velocity $v$, we choose $v\mathcal{T}=L_{\tau}=L$ such that the loop only has $L$ dependence. For the Wilson loop we have
\begin{align}
	W(L)&= \bra{G}S(L,\mathcal{T}/2)~~S(L,-\mathcal{T}/2) \ket{G} \notag \\
	&= \bra{G}S(L,\mathcal{T}/2)\tau^z_{s}(0) \tau^z_{s'}(0) \notag\\
	&\qquad\qquad~~\times\tau^z_{s}(0)  \tau^z_{s'}(0) S(L,-\mathcal{T}/2) \ket{G}\, .
\end{align}
We explicitly add and remove the charges for convenience. We define a state where we added a charge and anti-charge, that are a distance $L$ apart, to the ground state as 
\begin{equation}
	\ket{S(L)} = \tau^z_{s}(0)\tau^z_{s'}(0)~ S(L,-\mathcal{T}/2) \ket{G} \, .
\end{equation}
If we assume the distance to go to infinity, we can focus on the charge and neglect the anti-charge at infinity. Thus, the state $\ket{S(L\rightarrow \infty)}$ can be interpreted as a free charge state. Applying the reformulation to the (half) Wilson loop gives
\begin{equation}
	W_{1/2}(L) = \ip{G}{S(L)}\quad\text{and} \quad W(L) =\ip{S(L)}{S(L)}\, .
\end{equation}
The Fredenhagen-Marcu operator is rewritten to
\begin{equation}
	R(L) = \frac{W_{1/2}(L)}{\sqrt{W(L)}} = \frac{\ip{G}{S(L)}}{\sqrt{\ip{S(L)}{S(L)}}}\, ,
\end{equation}
\textit{i.e.}, the overlap between the ground state and the normalized charge anti-charge state. Taking the $L \rightarrow \infty$ limit indicates a phase containing free (deconfined) charge states $\ket{S(L\rightarrow \infty)}$ only if $R(L)$ decays to zero, since the state is orthogonal to the ground state. If in turn there is a finite overlap, the state $\ket{S(L)}$ would decay into the ground state signalling only confined charges~\cite{HuseFredMar, Fred_Mar_Quark}. Therefore, we write
\begin{equation}
	\lim\limits_{L \to \infty} R(L) = \begin{cases} 0\, , & \mathbb{Z}_2 \text{ deconfined,}\\
	\text{const.,} & \mathbb{Z}_2\text{ confined.} \end{cases} 
\end{equation}  
This diagnostic was first proposed in Ref.~\onlinecite{Fred_Mar_Quark} to test whether a theory contains free or confined quarks. It generally detects a confinement-deconfinement transition in a lattice gauge theory containing dynamical matter charges.

\section{Evaluation of the Fredenhagen-Marcu operator}

In this section, we discuss the perturbative evaluation of the Fredenhagen-Marcu operator in more detail.

\subsection{Dominant charge energy}\label{app:FM_calc_1}
For the regime of dominant charging energy, $E_C \gg E_J,E_M$, we found the unperturbed ground state in Eq.~\eqref{GS_EC}. For clarity, we expand the cosine functions and set $d_j=\text{e}^{-i \phi_j/2}$ to write the perturbation as
\begin{align}\label{pertbeiHC_eva}
	V=&- \frac{E_M}{2} \sum\limits_{<i,j>} \sigma_{ij}^z \left( d_{i}^{\dagger} d_{j}+ d_{i} d_{j}^{\dagger} \right) \notag\\
	&-\frac{E_J}{2} \sum\limits_{<i,j>} \left( {d_{i}^{\dagger}}^2 d_{j}^2 + d_{i}^2 {d_{j}^{\dagger}}^2 \right) \, .
\end{align}
The ground state to first order follows to
\begin{align}\label{pert_G_one}
	N' \ket{G'} &=\ket{G} + \frac{1}{4}\frac{E_M}{E_C} \sum\limits_{<i,j>} \sigma_{ij}^z \left( d_{i}^{\dagger} d_{j}+ d_{i} d_{j}^{\dagger} \right) \ket{G} \notag \\
	&\qquad + \frac{1}{16} \frac{E_J}{E_C} \sum\limits_{<i,j>} \left( {d_{i}^{\dagger}}^2 d_{j}^2 + d_{i}^2 {d_{j}^{\dagger}}^2 \right)\ket{G} \notag \\
	&= \ket{G} + \ket{G_1} + \ket{G_2} \, .
\end{align}
We define $N'$ to be the normalization factor. For the second term, the Majorana term, the two single charge creation/annihilation operators each lead to an $n_i^2= 1/4$ on the respective island and thus result in an energy change of $2 \cdot E_C$. This is the charging energy of two single charges. For the third term, the Josephson term, we analogously obtain $2 \cdot 4E_C$, the charging energy of two Cooper pairs. With this perturbative ground state one can calculate the expectation value of the (half) Wilson loop, both to first order. In the following, we denote the state with respect to which we calculate the expectation value of the Wilson loop explicitly by redefining $W(L) = \expval{W(L)}{G}$. As the mixed terms vanish, this yields
\begin{align}
	N'^2 &\expval{W(L)}{G'}  \notag\\
	&= \expval{W(L)}{G} +  \expval{W(L)}{G_1}+  \expval{W(L)}{G_2} \notag \\
	&= 1 + 2N  \left[ \frac{1}{4}\frac{E_M}{E_C}\right]^2  +2N  \left[ \frac{1}{16}\frac{E_J}{E_C} \right]^2  \, .
\end{align}  
The first term is the result for the unperturbed ground state. For the remaining terms it is important to note that the creation/annihilation operators $d_j^{\dagger}/d_j$ commute with the Pauli operators and that the perturbations are Hermitian. In the second term, the Majorana term, one finds that it only yields a finite contribution if the charge transfer occurs on the same link. This happens for $2 N$ terms of the sum, where $N$ is the number of nodes in the system. For the third term, the Josephson term, we analogously find those $2N$ summands, but with different prefactors. \\

Most importantly we note that the expectation value for the Wilson loop is finite to zeroth order already. If we now perform the same analysis for the half Wilson loop with a length $L$ larger than two links, we find a vanishing expectation value
\begin{align}
		N'^2 &\expval{W_{1/2}(L)}{G'} \notag \\
	&=  \left[ \frac{1}{4} \frac{E_M}{E_C} \right]^2  \sum\limits_{<i,j>}\sum\limits_{<k,l>}   \bra{G} \sigma_{ij}^z \left( d_{i}^{\dagger} d_{j}+ d_{i} d_{j}^{\dagger} \right)  \notag \\
	&\qquad\quad \times d_{s}^{\dagger}d_{s'}\left( \prod\nolimits_{r \in C_{1/2}}  \sigma^z_r \right) \sigma_{kl}^z \left( d_{k}^{\dagger} d_{l}+ d_{k} d_{l}^{\dagger} \right) \ket{G}\notag \\
	 &\qquad+ \left[ \frac{1}{16}\frac{E_J}{E_C} \right]^2  \sum\limits_{<i,j>}\sum\limits_{<k,l>}  \bra{G} \left( {d_{i}^{\dagger}}^2 d_{j}^2 + d_{i}^2 {d_{j}^{\dagger}}^2 \right) \notag \\
	&\qquad\quad \times d_{s}^{\dagger}d_{s'} \left( \prod\nolimits_{r \in C_{1/2}}  \sigma^z_r \right)  \left( {d_{k}^{\dagger}}^2 d_{l}^2 + d_{k}^2 {d_{l}^{\dagger}}^2 \right) \ket{G} \notag \\
	&= 0 \, . \label{half_Wilson_large_C}
\end{align}
From Eq.~\eqref{half_Wilson_large_C} we generalize to even higher orders. We find that contributions purely from the Josephson term cannot be finite even in higher orders of perturbation theory, since the Pauli operators $\sigma^z$ can never be squared to unity or form a closed contour. The first term yields a finite contribution if the open loop $C_{1/2}$ is bridged by consecutive single electron transfers. One can infer that even for mixed terms in higher orders, the lowest order contribution is still given by the pure single charge tunnelling. Those are at least of order $\mathcal{O} (\left[ E_M/(4E_C)\right]^{L})$. Now we are in a position to calculate the expectation value of the Fredenhagen-Marcu operator to leading order in the perturbation theory. With the ground state $\ket{G^{*}}$ and normalisation $N^{*}$ of at least $\order{L/2}$ in the perturbation, this reads
\begin{align}
	R(L) &= \frac{\expval{W_{1/2}(L)}{G^{*}}}{\sqrt{\expval{W(L)}{G^{*}}}N^{*}} \approx \frac{\left[ \frac{1}{4}\frac{E_M}{E_C} \right]^{L} }{1 \cdot 1}\notag\\
	&\approx \text{e}^{-\ln \left( 4\frac{E_C}{E_M}\right) L} \quad \overset{L \rightarrow \infty}{\longrightarrow} \quad 0 \, .
\end{align}
The Wilson loop as well as the normalization are approximated to leading order as $1$. Thus, the expectation value of the Fredenhagen-Marcu operator vanishes for $L \rightarrow \infty$ in the regime of dominant charging energy $E_C \gg E_J, E_M$.

\subsection{Dominant Josephson energy and large single-electron tunneling rate}\label{app:FM_calc_2}

For the case of dominant Josephson energy and large single-electron tunneling rate, $E_J \gg E_M \gg E_C$, we found the ground state in Eq.~\eqref{Gopen}. Furthermore, the perturbation is given by
\begin{equation}
    V = -\Delta \sum\limits_i \tau^x_i\, .
\end{equation}
We compute the first order perturbative ground state to
\begin{align}\label{pert_G_open}
	N' \ket{G'}&= \ket{G}  \underbrace{-\frac{1}{8}\frac{\Delta}{E_M}}_{=~\alpha} \sum\limits_{i} \tau_i^x \ket{G}\, .
\end{align}
We define the parameter $\alpha$ to simplify the notation. Since we are interested in the ground state expectation value of $R(L)$ for the limit of an infinite lattice we can, to a reasonable degree, approximate higher order corrections to the ground state as independent flips of matter DOF ($\tau^x_i$). Under this assumption of locally separate perturbations, we find for the ground state at order $N_{\text{max}}$
\begin{align}\label{pert_G_open_high}
	N^{*} \ket{G^{*}} &\approx \ket{G} + \sum\limits_{n=1}^{N_{\text{max}}} \frac{1}{n!} ~\alpha^n \Bigl( \sum\limits_{i} \tau_i^x \Bigr)^n \ket{G} \, .
\end{align}
At $n$-th order, we thus assume $n$ independent flips to occur, where the $n!$ is necessary to avoid double counting. The normalization is then calculated to
\begin{align}\label{normal}
	{N^{*}}^2 &=  \braket{G} + \bra{G}\sum\limits_{n=1}^{N_{\text{max}}} \frac{1}{n!} ~\alpha^n \Bigl( \sum\limits_{i} \tau_i^x \Bigr)^n \notag\\
	&\qquad \times\sum\limits_{m=1}^{N_{\text{max}}} \frac{1}{m!} ~\alpha^m \Bigl( \sum\limits_{j} \tau_j^x \Bigr)^m  \ket{G}\, .
\end{align}
Under the assumption of independent flips, the mixed order contributions have to vanish, as otherwise at least one $\tau_s^x$ does not square to unity. In particular, this means that the two sums in the calculation of the normalization in Eq.~\eqref{normal} simplify with a Kronecker delta $\delta_{nm}$ to
\begin{align}\label{normal_complete}
	{N^{*}}^2 &= 1 + \sum\limits_{n=1}^{N_{\text{max}}}  \frac{1}{n!} ~\alpha^{2n} \Bigl( \sum\limits_{i} \Bigr)^n \underbrace{\braket{G}}_{=~1} \notag \\
	&= \sum\limits_{n=0}^{N_{\text{max}}}  \frac{1}{n!} ~{(\alpha^{2} N)}^n ~\approx ~\text{e}^{\alpha^2 N}\, .
\end{align}
We used the fact that there are $n!$ ways to arrange the sums over $j$ such that all $\tau_i^x$ square to unity. The empty sums run over all nodes resulting in the total number of nodes $N$. In the last step, we added the remaining terms from $N_{\text{max}}$ to infinity, which is a negligible difference for large enough $N_{\text{max}}$. Moving on to the expectation value of the Wilson loop, we obtain to first order with Eq.~\eqref{pert_G_open}
\begin{align}
	N'^2 \expval{W(L)}{G'} &= 1 + \alpha^2 N\, .
\end{align}
Since the $\tau^x$ commute with the $\sigma^z$, we again obtain a Kronecker delta because the $\tau^x$ have to square to unity to result in a finite contribution. Thus, we repeat the above higher order calculation for the normalization factor for the Wilson loop to find
\begin{equation}\label{Wilson_approx}
	{N^{*}}^2 \expval{W(L)}{G^{*}}~ \approx ~ \text{e}^{\alpha^2 N}\, .
\end{equation}
In contrast to that, the half Wilson loop expectation value yields to first order 
\begin{align}
	N'^2 &\expval{W_{1/2}(L)}{G'}  \notag\\
&= 1 + \sum\limits_{i,j} \alpha^2 \bra{G} \tau_i^x~ \tau_s^z \tau_{s'}^z\prod\nolimits_{l \in \mathcal{C}_{1/2}} \sigma_l^z  ~\tau_j^x \ket{G} \notag  \\
	&= 1 + \alpha^2 \left[ (N-2) - 2 \right] \bra{G} \tau_s^z \tau_{s'}^z\prod\nolimits_{l \in \mathcal{C}_{1/2}} \sigma_l^z  \ket{G} \notag \\ 
	&= 1 + \alpha^2 (N-4)\, .
\end{align}
In total $N-2$ of the $\tau_j^x$ commute with $\tau^z_s$ and $\tau_{s'}^z$. In two terms the $\tau$ operators anticommute so that the expectation value differs from the Wilson loop. Assuming independent perturbations analogously to Eq.~\eqref{Wilson_approx} results in
\begin{equation}\label{half_Wilson_approx}
	{N^{*}}^2 \expval{W_{1/2}(L)}{G^{*}} ~\approx ~ \text{e}^{\alpha^2 (N-4)}\, .
\end{equation}  
Collecting Eq.~\eqref{normal_complete}, \eqref{Wilson_approx} and \eqref{half_Wilson_approx}, we obtain for the Fredenhagen-Marcu operator
\begin{align}
	R(L) &=  \frac{{N^{*}}^2 \expval{W_{1/2}(L)}{G^{*}}}{{N^{*}}\sqrt{{N^{*}}^2\expval{W(L)}{G^{*}}}} = \text{e}^{-4\alpha^2 }\, .
\end{align}
Resubstitution of the parameter gives
\begin{equation}
	 R(L) = \text{e}^{-\frac{1}{16}{\left( \frac{\Delta}{E_M}\right)}^2} \quad\overset{L \rightarrow \infty}{\longrightarrow}\quad \text{const.}
\end{equation}
Thus, for $E_C \gg E_M \gg E_C$ the Fredenhagen-Marcu operator takes a constant value and the phase is confining.

\section{Perturbation analysis\label{sec:perturbdescr}}

In this section we give a more detailed description of the perturbation analysis for the case of $E_C \gg E_M,E_J$. Let us start by giving some additional notation. We define the projectors
\begin{equation}
	P_- = \sum\limits_{G}\dyad{G}\qquad\text{and} \qquad P_{+} = 1 - P_- \, .
\end{equation}
The $P_-$ projects onto the ground state space and $P_+$ projects onto the remaining space. With
\begin{equation}
	G_{0+}(E) = P_+\frac{1}{E-H_{0}} P_+ \, ,
\end{equation} 
$V_{\pm \mp} = P_{\pm} V P_{\mp}\qquad \text{and} \qquad H_{0+} = P_+ H_0 P_+$, the effective Hamiltonian is given by
\begin{equation}\label{BWHam}
	H_{\text{eff}} = V_{-+} G_{0+}(E) \sum_{k=0}^{\infty} \left( V_{++} G_{0+}(E)\right)^k V_{+-} \, .
\end{equation} 
We neglected the constant contribution and linear corrections vanish. Note that $\ket{G}$ is in general highly degenerate, if we only take $H_0$ into account. In the presence of the perturbation, effectively introducing the plaquette terms and the gauge constraint, this degeneracy is lifted for the most part. The ground state then remains $4$-fold degenerate due to the topology of the system under consideration. Switching between these four states will only be possible in $L_{\text{min}}/2$-th order, where $L_{\text{min}}$ denotes the total length of the smallest lattice dimension. Thus, for reasonably low orders of perturbation theory there is no mixing and we can concentrate on one of the ground states. It was proven, that the ground states remain degenerate even in the perturbed case~\cite{HuseFredMar}. To do this, they defined two ground states $\ket{G}$ and $\ket{G'}=\Gamma \ket{G}$, where  $\Gamma = \prod_{l \in \mathcal{C}_{nc}} \sigma^z_l$ denotes a non-contractible loop. Then one can show that
\begin{align}
	\bra{G} V P_+ G_{0+} \dots P_+ V \ket{G} = \bra{G'} V P_+ G_{0+} \dots P_+ V \ket{G'}
\end{align}
by expressing $\ket{G'}$ in terms of $\ket{G}$ and commuting the latter $\Gamma$ to the front such that it squares to identity. Therefore, we only perform the calculation for one representative ground state. \\
Since the Green's function for the ground state energy corrections is negative, we define $\tilde{G}_{0+}(E) = -G_{0+}(E)$ and $\tilde{V}=-V$ to find
\begin{equation}
    H_{\text{eff}} = -\tilde{V}_{-+} \tilde{G}_{0+}(E) \sum_{k=0}^{\infty} \left( \tilde{V}_{++} 
    \tilde{G}_{0+}(E)\right)^k \tilde{V}_{+-} \, .
\end{equation}
This allows us to keep better track of the signs and we omit the $\sim$ again for convenience. In contrast to earlier works, where the perturbation was only performed up to fourth order, we have to expand $G_{0+}$ also in terms of $E$ around $E_0=0$. We can compute this with the operator identity
\begin{equation}
	\frac{1}{X-Y} = \frac{1}{X} + \frac{1}{X}Y\frac{1}{X-Y}
\end{equation}
as 
\begin{align}
	G_{0+}(E) &= \frac{1}{H_{0+}-E} = \frac{1}{H_{0+}-(0+\delta E)} \notag\\
	&= \frac{1}{H_{0+}} + \frac{1}{H_{0+}}~ \delta E~ \frac{1}{H_{0+}-\delta E} \notag \\
	&=\frac{1}{H_{0+}} + \frac{1}{H_{0+}}~ \delta E~\frac{1}{H_{0+}} \notag\\
	&~~~~+  \frac{1}{H_{0+}}~ \delta E~ \frac{1}{H_{0+}}~ \delta E~\frac{1}{H_{0+}} + \order{V^6} \, .
\end{align}

To capture all important terms up to sixth order, we now have to set
\begin{equation}
	\delta E =\delta E^{(2)} + \delta E^{(3)} +\delta E^{(4)} + \order{V^5} \, ,
\end{equation}
since there is no linear contribution. The third order is furthermore neglected as it cannot lead to a plaquette contribution up to sixth order. The expansion has to be inserted back into Eq.~\eqref{BWHam} to find all the necessary terms. Collecting the results, we find
\begin{align}\label{Heffsum}
	H_{\text{eff}} &= \text{const.} + H_{\text{eff}}^{(4)} + H_{\text{eff}}^{(5)} + H_{\text{eff}}^{(6)} \, ,
\end{align}
where
\begin{align}
     H_{\text{eff}}^{(4)} &= -V_{-+} G_{0+}  \left[ V_{++} G_{0+} \right]^2 V_{+-} \notag \\
     H_{\text{eff}}^{(5)} &= -V_{-+} G_{0+}  \left[ V_{++} G_{0+} \right]^3 V_{+-} \notag \\
     H_{\text{eff}}^{(6)} &= -V_{-+} G_{0+}  \left[ V_{++} G_{0+} \right]^4 V_{+-} \notag\\
     &~~~-  \delta E^{(2)} V_{-+} G_{0+} M^2  G_{0+} V_{+-} \notag\\
     &~~~- \delta E^{(4)}V_{-+} G_{0+}^2 V_{+-}
\end{align}
and
\begin{equation}
	M^2 = 2V_{++} G_{0+} V_{++} G_{0+} + V_{++}  G_{0+}^2 V_{++} \, .
\end{equation}
Here, we set $G_{0+} = 1/H_{0+}$. Without the energy expansion, we would obtain only negative contributions to the effective Hamiltonian.

\bibliography{Supercond}

\end{document}